\documentclass[aps,pre,showpacs,onecolumn,a4paper,floatfix,superscriptaddress]{revtex4}
\usepackage{amssymb}
\usepackage{amsfonts}
 \usepackage{amsmath}
\usepackage{graphics}
\usepackage{times}
\usepackage{color}
\usepackage{epsfig}

\newcommand{\be}{\begin{equation}}
\newcommand{\ee}{\end{equation}}

\newcommand{\bea}{\begin{eqnarray}}
\newcommand{\eea}{\end{eqnarray}}

\begin{document}

\title{Dynamics of interacting diseases}
\author{Joaqu\'in Sanz}
 \affiliation{Institute for Biocomputation and Physics of Complex Systems (BIFI), University of Zaragoza, Zaragoza 50018, Spain}
 \affiliation{Department of Theoretical Physics, University of Zaragoza, Zaragoza 50009, Spain}

\author{Cheng-Yi Xia}

 \affiliation{Institute for Biocomputation and Physics of Complex Systems (BIFI), University of Zaragoza, Zaragoza 50018, Spain}
\affiliation{Key Laboratory of Computer Vision and System (Ministry of Education) and Tianjin Key Laboratory of Intelligence Computing and Novel Software Technology, Tianjin University of Technology, Tianjin 300191, P.R. China}

\author{Sandro Meloni}
 \affiliation{Institute for Biocomputation and Physics of Complex Systems (BIFI), University of Zaragoza, Zaragoza 50018, Spain}

\author{Yamir Moreno}
 \affiliation{Institute for Biocomputation and Physics of Complex Systems (BIFI), University of Zaragoza, Zaragoza 50018, Spain}
 \affiliation{Department of Theoretical Physics, University of Zaragoza, Zaragoza 50009, Spain}
\affiliation{Complex Networks and Systems Lagrange Lab, Institute for Scientific Interchange, Turin, Italy}

\begin{abstract}  

Current modeling of infectious diseases allows for the study of complex and realistic scenarios that go from the population to the individual level of description. However, most epidemic models assume that the spreading process takes place on a single level (be it a single population, a meta-population system or a network of contacts). 
In particular, interdependent contagion phenomena can only be addressed if we go beyond the scheme one pathogen-one network. In this paper, we propose a framework that allows describing the spreading dynamics of two concurrent  diseases. Specifically, we characterize analytically the epidemic thresholds of the two diseases for different scenarios and also compute the temporal evolution characterizing the unfolding dynamics. Results show that there are regions of the parameter space in which the onset of a disease's outbreak is conditioned to the prevalence levels of the other disease. Moreover, we show, for the SIS scheme, that under certain circumstances, finite and not vanishing epidemic thresholds are found even at the thermodynamic limit for scale-free networks. For the SIR scenario, the phenomenology is richer and additional interdependencies show up. We also find that the secondary thresholds for the SIS and SIR models are different, which results directly from the interaction between both diseases. Our work thus solve an important problem and pave the way towards a more comprehensive description of the dynamics of interacting diseases.

 \end{abstract}
 

\maketitle


\newpage


\section{Introduction}

The problem of how diseases spread has been studied for a long time \cite{Maybook,hethcote,daley,Murray,Marro,allen,stanley,heth76}. In the last few years, due to the increasing availability of data about transmission patterns, contact networks and population mobility, it has become apparent that we are in the position to develop theoretical and computational frameworks that will ultimately allow the forecast of epidemic outbreaks \cite{vespiromu,vespbook08,llm01,yamirrep,dgm08,newmanrev,geisel,egamstw04,vrpre,n02,mpsv02,caste,bogupas,bbpsv04,marian,Guerra,EPLnostre,glmp08}. The latter is a consequence of the increasing availability of new models that are capable of providing better predictions of real epidemic scenarios \cite{Gleam}. In particular, we have been able to identify what ingredients are key to characterize the unfolding of a disease outbreak. This is the case of the structure of the population in the form of networks of contacts at a local scale \cite{vespiromu,vrpre,mpsv02,sfm10}, or of individual mobility patterns that facilitate the spreading of diseases in wider geographical areas \cite{cpv07}. The theoretical implications of considering these aspects have been thoroughly addressed \cite{meloni1,meloni2,bgb11,poletto}, and at the same time, the amount and quality of real data that are relevant to epidemic spreading is constantly increasing \cite{stanley_nat,polymod}. 

In this context, an emergent field of research is the modeling of coupled spreading phenomena, whether they are two pathogens or multiple strains of the same disease that propagate concurrently on the same population \cite{poletto,kn11,fj10,Newman-old,Marceau}. Focusing on the two-pathogen scenario, the complexity of the problem increases because now the natural history of one of the diseases is affected by the presence of the second one, typically as a consequence of the modification of the host's immune response after infection $-$ with a plethora of possibilities as given by different interaction schemes $-$. In addition, the networks of contacts through which the pathogens spread can vary from one disease to the other. Typical examples of these coupled spreading phenomena are given by the interaction between HIV infection and the spreading of certain opportunist  pathogens \cite{SIDA-virus,SIDA-bacterias}, or the strain-strain interactions of a viral pathogen like the flu virus \cite{flu-flu}. 

From a theoretical point of view, one of the first works that considered the above problem from a networked point of view is due to Funk and Jensen in 2010 \cite{fj10}. In their work, the authors exploited the analogy of the spreading process with a bond-percolation problem so as to address the issue of epidemic thresholds in a SIR (Susceptible-Infected-Removed) model. However, this first work did not consider a framework in which the temporal evolution of the epidemics can be studied. Following a similar bond-percolation approach Mills et al. \cite{mills2013}, motivated by the interaction between AIDS and Tuberculosis, studied two coupled SIR models where the diseases can spread in both layers at different rates. Also in this case, due to the {\it static} nature of the bond-percolation approach, the temporal evolution of the diseases cannot be studied. Recently, Marceau et al. presented a new model \cite{Marceau} aimed at studying the latter aspect, also in within SIR scheme, using ``on-the-fly graphs'', a network generation model previously introduced by the authors \cite{on-the-fly}. In this work, although the temporal evolution of the system can be monitored, the modeling approach does not provide information about epidemic thresholds. In addition, the model can not be trivially generalized so as to cover other classical models like SIS (Susceptible-Infected-Susceptible). Other works have considered the problem by studying the spreading of the diseases as a Markovian process. In \cite{sahneh2} the authors, using two coupled Markov processes, studied the {\em competition} of two epidemics and found different regions of the parameters space where the two diseases can coexist. However, the model in \cite{sahneh2} only represents the specific scenario of two mutually excluding diseases. It is also worth mentioning that using a similar approach, the so-called microscopic Markov-Chain \cite{EPLnostre}, the specific case in which a disease and an "awareness" diffusion process coexist has been studied in \cite{granell2013}. 

In this work, we propose a modeling framework, {\color{black}based on an heterogeneous mean field (HMF) approximation}, for the spreading of two concurrent diseases that interact with each other. To this purpose, we analyze in detail the dynamics of a two-layer system that represents the networks of contacts on top of which either two coupled SIS or SIR processes spread. Although we use as a convenient representation of the system under study a topology that has attracted a lot of attention lately, i.e., the so-called multilayer networks \cite{mikko_review,PRX}, we stress that the main focus is to study what are the effects of the complex interplay between the different interaction mechanisms considered and the temporal and topological scales onto the dynamical behavior of the diseases. Specifically, we explicitly derive the epidemic threshold of each disease in terms of the parameters that characterize their evolution and the topology of the networks of contacts. Remarkably, we find that the epidemic thresholds in the SIR case are different from those of the SIS case, which is a consequence of the emergence of new mechanisms of interaction and to the transitory nature of the epidemic outbreaks in the SIR case. Additionally, different scenarios and limiting cases of both theoretical and epidemiological interest are scrutinized. Finally, results from numerical simulations are presented to validate our analytical results. Our work thus shows how the different interaction mechanisms considered can give rise to new phenomenological insights regarding the dynamics of interacting diseases.

\section{Modeling Framework}

Our purpose in this work is to explore epidemiological scenarios in which two different infectious diseases simultaneously spread through the same host population, and whose dynamical parameters (i.e. infectiousnesses and recovery rates) depend, at the level of single individuals, on whether the agents involved have caught only one of the diseases or both. In addition, we consider that neither the mechanisms behind each disease evolution nor the contact networks through which they spread have to be the same, and thus, are considered independently for each disease. Let us then consider that we have two diseases (disease $1$ and disease $2$) which spread over two different networks of contacts: disease $1$ propagates over network $1$, which has a mean degree $\langle k\rangle=\sum_{{k,l}}{P(k,l)k}$; whilst disease $2$ does so over network $2$, whose mean connectivity is equal to $\langle l\rangle=\sum_{{k,l}}{P(k,l)l}$. The composed degree distribution $P(k,l)$ gives the proportion of nodes (individuals) having $k$ and $l$ links in networks 1 and 2, respectively \cite{note1}. In what follows, we consider independently both the SIS and SIR schemes.
\begin{figure}
\begin{center}
  \includegraphics[width=\columnwidth]{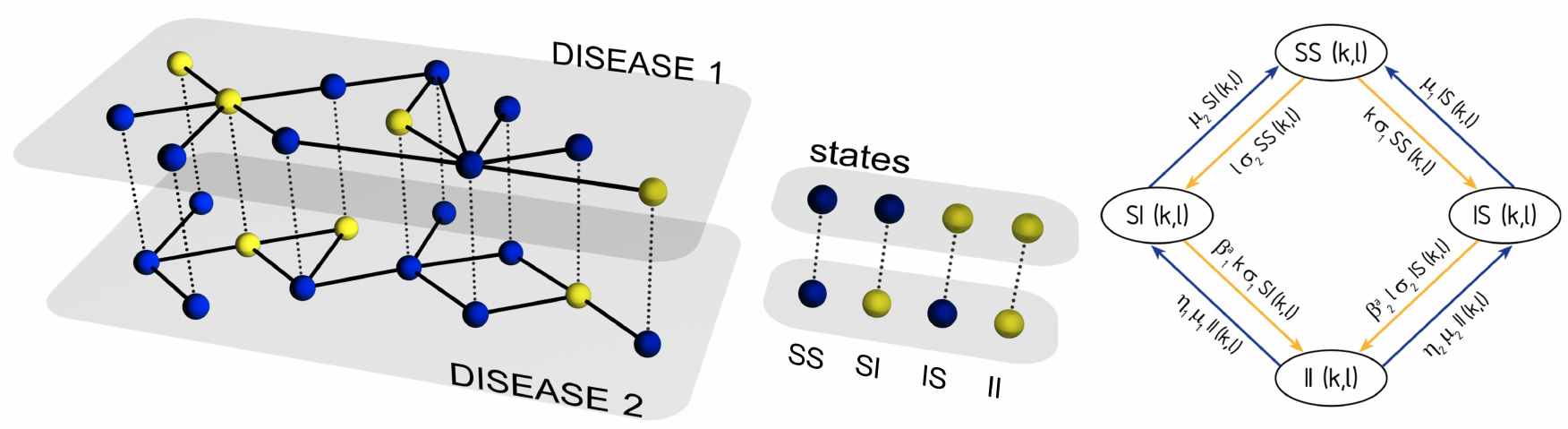}
\caption{(color online) SIS-SIS interacting diseases model (a) Each of the diseases spreads over a different network. Each individual belongs simultaneously to each network and can be (or not) infected with each of the diseases. (b) Set of transitions allowed in the model. The variables $(SS(k,l),IS(k,l),SI(k,l),II(k,l))$ represent the densities of individuals of each type in the system having $k$ neighbors in network 1 and $l$ neighbors in network $2$.}
\label{figure1}
\end{center}
\end{figure}

\subsection{The SIS scenario}

As a first step, we will consider the baseline scenario in which the isolated dynamics of each disease, when the second is absent, is described by a simple S-I-S scheme. Each individual belonging to a composed connectivity class $(k,l)$ can be in four different dynamical states: susceptible with respect to both diseases $SS(k,l)$, infected of both $II(k,l)$, and infected with the first (second) one and still susceptible to catch the second (first) disease, $IS(k,l)$ ($SI(k,l)$), being these quantities the proportion of individuals at each disease state with composed degree class $(k,l)$. Thus, we have that $SS(k,l)+IS(k,l)+SI(k,l)+II(k,l)=1$ $\forall(k,l)$. In addition, regardless of the connectivities of the nodes involved, we have eight possible contagion transitions after a contact (four for each disease). On the other hand, four other possible transitions correspond to the cases in which infected individuals go back to the susceptible class. This amounts to a total of twelve elementary transitions, schematized as follows (see also Figure.\ \ref{figure1}): 

\begin{center}
\begin{eqnarray*}
IS+SS\xrightarrow{\lambda_{1}}IS+IS\\
IS+SI\xrightarrow{\beta_{1}^{a}\lambda_{1}}IS+II\\
II+SS\xrightarrow{\beta_{1}^{b}\lambda_{1}}II+IS\\
II+SI\xrightarrow{\beta_{1}^{a}\beta_{1}^{b}\lambda_{1}}II+II\\
SI+SS\xrightarrow{\lambda_{2}}SI+SI\\
SI+IS\xrightarrow{\beta_{2}^{a}\lambda_{2}}SI+II\\
II+SS\xrightarrow{\beta_{2}^{b}\lambda_{2}}II+SI\\
II+IS\xrightarrow{\beta_{2}^{a}\beta_{2}^{b}\lambda_{2}}II+II\\
IS\xrightarrow{\mu_{1}}SS\\
SI\xrightarrow{\mu_{2}}SS\\
II\xrightarrow{\eta_{1}\mu_{1}}SI\\
II\xrightarrow{\eta_{2}\mu_{2}}IS
\end{eqnarray*}
\end{center}

\begin{table}
\begin{center}
\begin{tabular}{c|p{87mm}l}
\hline \hline
Parameter & Dynamical meaning \\\hline
$\lambda_{1}$& Baseline infectiousness of disease $1$ \\\hline
$\lambda_{2}$& Baseline infectiousness of disease $2$ \\\hline
$\mu_{1}$& Baseline recovery rate of disease $1$ \\\hline
$\mu_{2}$& Baseline recovery rate of disease $2$ \\\hline
$\beta_{1}^{a}$ & Variation of disease $1$ infectiousness due to the fact that the susceptible individual exposed to disease $1$ is infected with disease $2$\\\hline
$\beta_{2}^{a}$ & Variation of disease $2$ infectiousness due to the fact that the susceptible individual exposed to disease $2$ is infected with disease $1$\\\hline
$\beta_{1}^{b}$ & Variation of disease $1$ infectiousness due to the fact that the spreader is also infected with disease $2$\\\hline
$\beta_{2}^{b}$ & Variation of disease $2$ infectiousness due to the fact that the spreader is also infected with disease $1$\\\hline
$\eta_{1}$ & Variation of disease $1$ recovery rate for individuals also infected with disease $2$\\\hline
$\eta_{2}$ & Variation of disease $2$ recovery rate for individuals also infected with disease $1$\\\hline\hline
\end{tabular}
\caption{Definition of model parameters}
\label{table1}
\end{center}
\end{table}

The model thus contains two basic infection probabilities --$\lambda_{1}$ and $\lambda_{2}$-- as well as two basic recovery rates --$\mu_{1}$ and $\mu_{2}$--, one for each disease. In addition, infection probabilities are affected by scaling factors --the four $\beta$'s and combinations of them-- and so are the recovery rates --by the $\eta$'s--, as it is explained in detail in Table \ref{table1}. These parameters describe the interaction of the diseases through three different effects that are concurrently taken into account. The first effect is the variation of the susceptibility of healthy individuals to get infected with one disease as a consequence of being infected with another. This mechanism is described by $\beta_{1}^{a}$ for the variation of the infection risk of disease 1 caused by disease 2, and $\beta_{2}^{a}$, for the symmetric case. The second effect is the change of the spreading capabilities of double-infected individuals with respect to single-infected ones, which is described by the parameters $\beta_{1}^{b}$ and $\beta_{2}^{b}$, for diseases 1 and 2, respectively. The last effect is the variation of the infectious periods of double-infected individuals also with respect to single-infected ones, described by $\eta_{1}$ and $\eta_{2}$.

Therefore, these parameters exhaustively describe all the ways in which two diseases can interact according to a SIS scheme, and allow us to isolate the different effects of one disease on the spreading of the other by making infectious individuals more efficient spreaders or by making susceptible individuals more prone to get sick. Once we have defined the whole set of parameters in table \ref{table1}, we have all possible transitions between dynamical states well defined (see Figure \ref{figure1}). 

According to the scheme depicted in the figure (in which simultaneous double contagions and recoveries from both diseases have been explicitly excluded), {\color{black}we can consider that all individuals within the same composed connectivity class $(k,l)$ are dynamically equivalent, in order to get a composed HMF for the dynamical description of both diseases which neglects, as a first approximation, further correlations}. In this way, the set of differential equations describing the evolution in time of the four densities of individuals $(SS(k,l),IS(k,l),SI(k,l),II(k,l))$ is as follows:

\begin{eqnarray}
\dot{SS}(k,l)=-k\lambda_{1} \theta_{1}^{IS}SS(k,l)-l\lambda_{2}\theta_{2}^{SI}SS(k,l)-k\beta_{1}^{b}\lambda_{1}\theta_{1}^{II}SS(k,l)-l\beta_{2}^{b}\lambda_{2}\theta_{2}^{II}SS(k,l)+\mu_{1}IS(k,l)+\mu_{2}SI(k,l)\\
\dot{IS}(k,l)=k\lambda_{1}\theta_{1}^{IS}SS(k,l)+k\beta_{1}^{b}\lambda_{1}\theta_{1}^{II}SS(k,l)-l\beta_{2}^{a}\lambda_{2}\theta_{2}^{SI}IS(k,l)-l\beta_{2}^{a}\beta_{2}^{b}\lambda_{2}\theta_{2}^{II}IS(k,l)-\mu_{1}IS(k,l)+\eta_{2}\mu_{2}II(k,l)\\
\dot{SI}(k,l)=l\lambda_{2}\theta_{2}^{SI}SS(k,l)+l\beta_{2}^{b}\lambda_{2}\theta_{2}^{II}SS(k,l)-k\beta_{1}^{a}\lambda_{1}\theta_{1}^{IS}SI(k,l)-k\beta_{1}^{a}\beta_{1}^{b}\lambda_{1}\theta_{1}^{II}SI(k,l)-\mu_{2}SI(k,l)+\eta_{1}\mu_{1}II(k,l)\\
\dot{II}(k,l)=k\beta_{1}^{a}\lambda_{1}\theta_{1}^{IS}SI(k,l)+l\beta_{2}^{a}\lambda_{2}\theta_{2}^{SI}IS(k,l)+k\beta_{1}^{a}\beta_{1}^{b}\lambda_{1}\theta_{1}^{II}SI(k,l)+l\beta_{2}^{a}\beta_{2}^{b}\lambda_{2}\theta_{2}^{II}IS(k,l)-(\eta_{1}\mu_{1}+\eta_{2}\mu_{2})II(k,l)
\end{eqnarray}
where the $\theta$ parameters are defined in Table \ref{table2}.

\begin{table}
\begin{center}
\begin{tabular}{c|c}
\hline\hline
$\theta_{1}^{IS}=\frac{\sum_{k,l}{P(k,l)kIS(k,l)}}{\sum_{k,l}{P(k,l)k}}$ & probability that a given link of network 1 points to an $IS$ node \\\hline
$\theta_{1}^{II}=\frac{\sum_{k,l}{P(k,l)kII(k,l)}}{\sum_{k,l}{P(k,l)k}}$ & probability that a given link of network 1 points to an $II$ node \\\hline
$\theta_{2}^{SI}=\frac{\sum_{k,l}{P(k,l)lSI(k,l)}}{\sum_{k,l}{P(k,l)l}}$ & probability that a given link of network 2 points to a $SI$ node \\\hline
$\theta_{2}^{II}=\frac{\sum_{k,l}{P(k,l)lII(k,l)}}{\sum_{k,l}{P(k,l)l}}$ & probability that a given link of network 2 points to an $II$ node \\\hline\hline
\end{tabular}
\caption{Definition of parameters $\theta$.}
\label{table2}
\end{center}
\end{table}

\begin{figure}
  \begin{center}
  \includegraphics[]{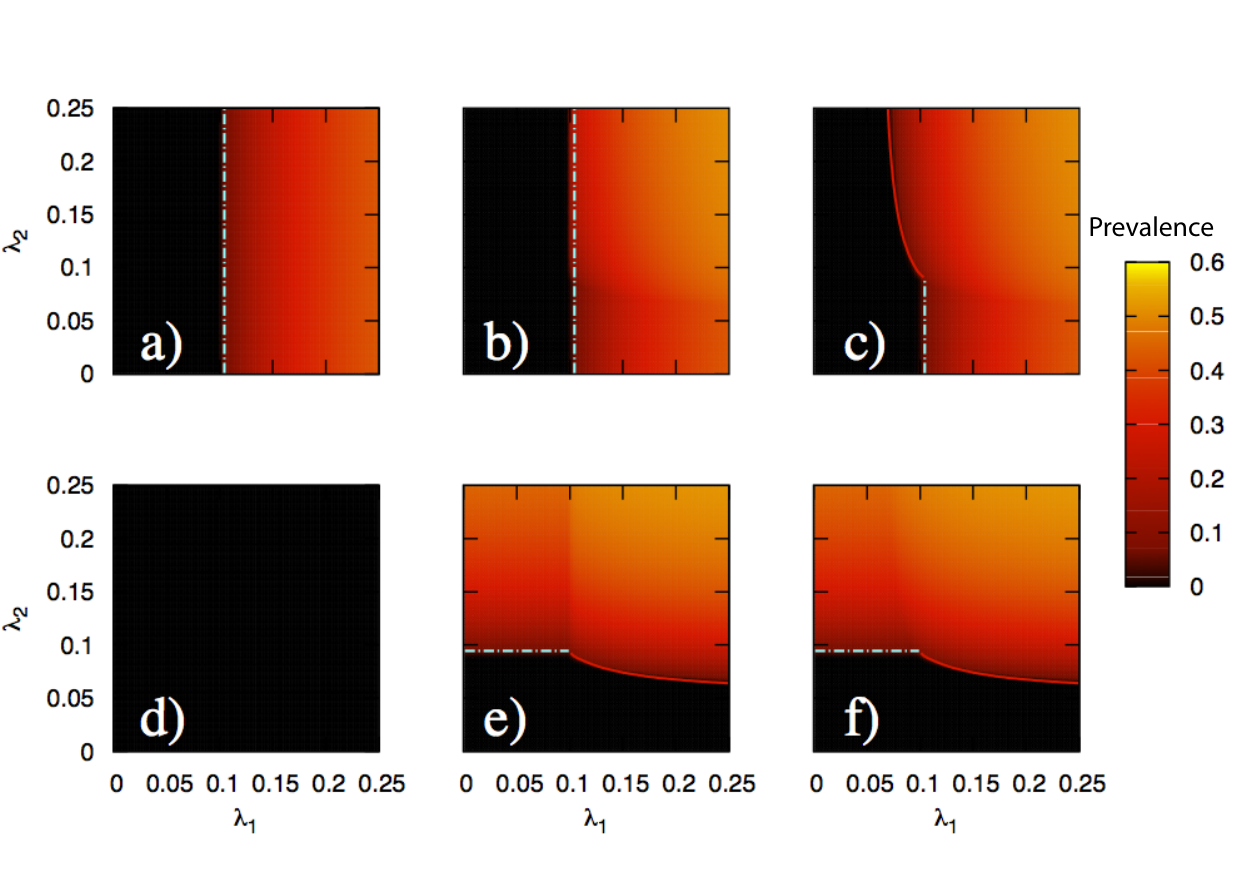}
  \end{center}
\caption{(color online) Reciprocally enhanced spreading. Dynamical parameters: $\mu_{1}=\mu_{2}=0.75$, $\beta_{1}^{a}=\beta_{1}^{b}=\beta_{2}^{a}=\beta_{2}^{b}=1.3$, $\eta_{1}=\eta_{2}=0.8$. Networks: Erd\"os-Renyi graphs: $N_{1}=N_{2}=5000$ agents, $\langle k\rangle=7$,  $\langle l\rangle=8$. The color maps represent the prevalence levels of diseases 1 -- upper panels -- and 2 -- lower panels --, at different stages of the Monte-Carlo simulations. Step 1 (panels a,d): stationary levels after the initial introduction of an infection seed of disease 1 ($\epsilon_{IS}=0.005$). Step 2 (panels b,e): once stage 1 is completed, an infection seed of disease 2 ($\epsilon_{SI}=0.005$) is introduced, and stationarity is recovered. Step 3: (panels c,f): after stage 2, an additional seed of infection 1 is re-introduced ($\epsilon_{IS}=0.005$), and the stationary prevalences plotted. Dashed and solid lines represent respectively primary and secondary thresholds predicted by the model.}
\label{fig2}
\end{figure}

Combining the probabilities $\theta$s and defining the following two new parameters $\sigma_{1}=\lambda_{1}(\theta_{1}^{IS}+\beta_{1}^{b}\theta_{1}^{II})$ and $\sigma_{2}=\lambda_{2}(\theta_{2}^{SI}+\beta_{2}^{b}\theta_{2}^{II})$ we obtain the average probabilities per link for {\color{black}$SS$}-nodes to become infected with disease 1 ($\sigma_{1}$) or disease 2 ($\sigma_{2}$). This allows us to rewrite the system of differential equations as:
\begin{eqnarray}
\dot{SS}(k,l)=-(k\sigma_{1}+l\sigma_{2})SS(k,l)+\mu_{1}IS(k,l)+\mu_{2}SI(k,l)\\
\dot{IS}(k,l)=k\sigma_{1}SS(k,l)-l\beta_{2}^{a}\sigma_{2}IS(k,l)-\mu_{1}IS(k,l)+\eta_{2}\mu_{2}II(k,l)\\
\dot{SI}(k,l)=l\sigma_{2}SS(k,l)-k\beta_{1}^{a}\sigma_{1}SI(k,l)-\mu_{2}SI(k,l)+\eta_{1}\mu_{1}II(k,l)\\
\dot{II}(k,l)=k\beta_{1}^{a}\sigma_{1}SI(k,l)+l\beta_{2}^{a}\sigma_{2}IS(k,l)-(\eta_{1}\mu_{1}+\eta_{2}\mu_{2})II(k,l)
\label{sistema}
\end{eqnarray}
%
Due to the closure relationship $SS(k,l)+IS(k,l)+SI(k,l)+II(k,l)=1$ $\forall (k,l)$, only three of these four equations are linearly independent for each composed connectivity class $(k,l)$. Taking this into account, we next analyze the time evolution of the vector $(IS(k,l),SI(k,l),II(k,l))$, since $SS(k,l)=1-IS(k,l)-SI(k,l)-II(k,l)$.

\subsubsection{Epidemic thresholds}

In order to analyze the most relevant dynamical properties of the system, we look for the values $(IS^{*}(k,l),SI^{*}(k,l),II^{*}(k,l))$ that define the stationary state: $(\dot IS^{*}(k,l),\dot SI^{*}(k,l),\dot II^{*}(k,l))=(0,0,0)$ $\forall(k,l)$ for the system Eq. (\ref{sistema}). In order to do that, we are forced to consider $\sigma_{1}$ and $\sigma_{2}$ as additional parameters, although these two quantities are linear combinations of the other variables and the ultimate responsible of the coupling among all connectivity classes. As it happens commonly in this kind of models \cite{vrpre,allen}, there exists a trivial fixed point of the dynamics in which there are no infected individuals in the system: $(II^{*}_{1}(k,l),II^{*}_{2}(k,l),II^{*}(k,l))=(0,0,0)$ $\forall (k,l)$. This fixed point  represents the absorbing state of our model. In addition, there are other possible fixed points for which the densities of infected individuals could be written as a function of the $\sigma$ parameters:  $(II^{*}_{1}(k,l,\sigma_{1},\sigma_{2}),II^{*}_{2}(k,l,\sigma_{1},\sigma_{2}),II^{*}(k,l,\sigma_{1},\sigma_{2}))\neq(0,0,0)$. 

It is thus possible to get self-consistent equations for the variables $\sigma$ as:
\begin{eqnarray}
\sigma_{1}=f_{1}(\sigma_{1},\sigma_{2})=\frac{\lambda_{1}}{\langle k \rangle}\sum_{k,l}{P(k,l)k(IS(k,l,\sigma_{1},\sigma_{2})+\beta_{1}^{b}II(k,l,\sigma_{1},\sigma_{2})})\\
\sigma_{2}=f_{2}(\sigma_{1},\sigma_{2})=\frac{\lambda_{2}}{\langle l \rangle}\sum_{k,l}{P(k,l)l(SI(k,l,\sigma_{1},\sigma_{2})+\beta_{2}^{b}II(k,l,\sigma_{1},\sigma_{2})})
\label{trascen}
\end{eqnarray}
The condition $\sigma_{1}=f_{1}(\sigma_{1},\sigma_{2})>0$ implies the existence of a stable active state for the dynamics of disease 1, i.e., a state in which disease 1 becomes endemic in the population. For this situation to take place for disease 2, the condition $\sigma_{2}=f_{2}(\sigma_{1},\sigma_{2})>0$ must be fulfilled.
%
\begin{figure}
  \begin{center}
  \includegraphics[]{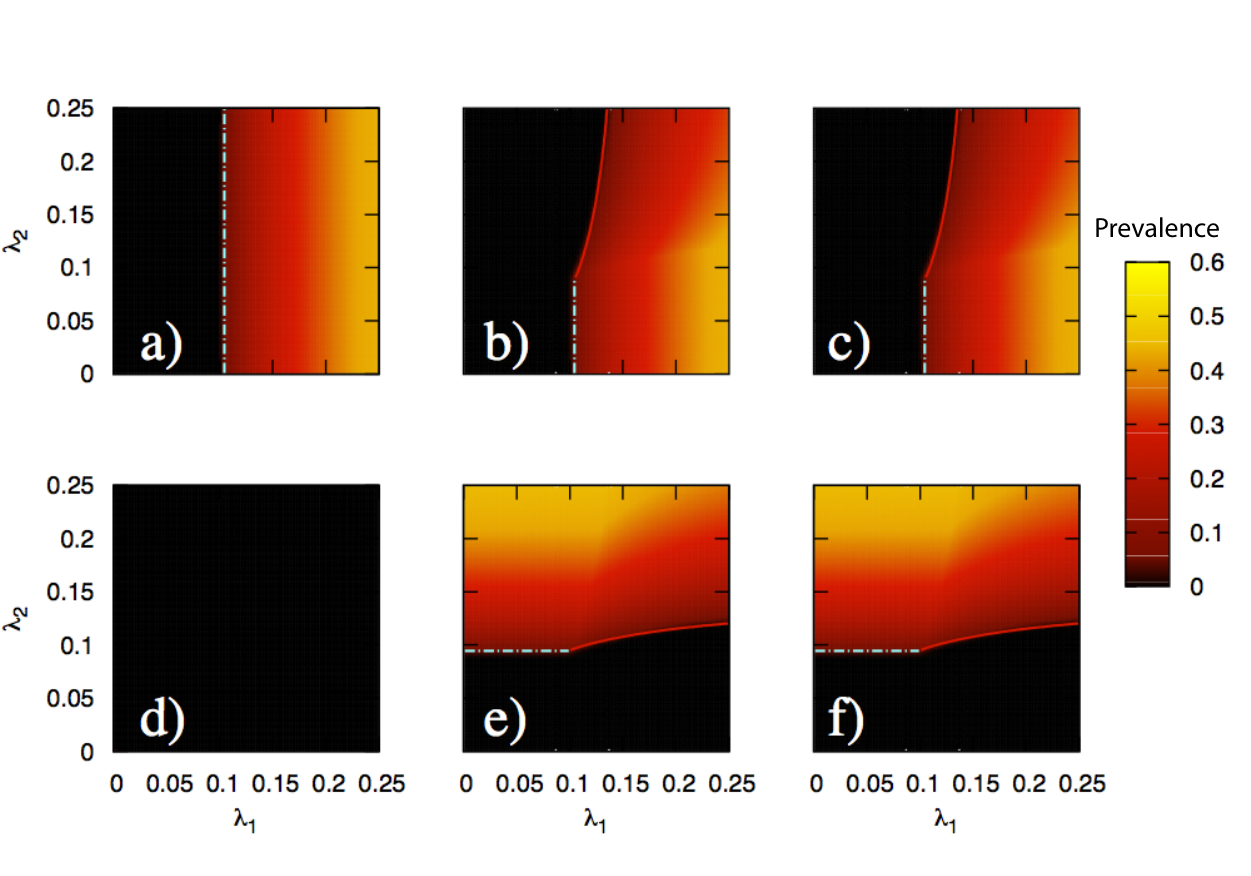}
  \end{center}
\caption{(color online) Reciprocally impaired spreading. Dynamical parameters: $\mu_{1}=\mu_{2}=0.75$, $\beta_{1}^{a}=\beta_{1}^{b}=\beta_{2}^{a}=\beta_{2}^{b}=0.8$, $\eta_{1}=\eta_{2}=1.3$. Networks: Erd\"os-Renyi graphs: $N_{1}=N_{2}=5000$ agents, $\langle k\rangle=7$,  $\langle l\rangle=8$. The color maps represent the prevalence levels of diseases 1 -- upper panels -- and 2 -- lower panels --, at different stages of the Monte-Carlo simulations. As it is done in figure \ref{fig2}, we introduce successively three infectious seeds $(IS,SI,IS)=(0.005,0.005,0.005)$, and plot the stationary prevalences after each fluctuation in the three columns of the figure. As it can be seen, the reintroduction of the third seed of infection 1 in the system does not affect the prevalence levels, as global stability is reached after the second stage.}
\label{fig3}
\end{figure}
%
Given the symmetry between the two expressions in Eq. (\ref{trascen}), we will only focus on the analysis of the first equation. In fact, it can be shown that always $\frac{{\partial^{2} f(\sigma_{1},\sigma_{2})}}{{\partial \sigma_{1}^{2}}}<0$, which means that, if we think of the graphical solution of the equation $\sigma_{1}=f_{1}(\sigma_{1},\sigma_{2})$, for this non trivial solution to exist --given that, obviously $f(\sigma_{1}=0,\sigma_{2}=0)=0$-- it must be verified that $\left[\frac{{\partial f(\sigma_{1},\sigma_{2})}}{{\partial \sigma_{1}}}\right]_{\sigma_{1}=0}>1$, as in \cite{vrpre}. After some algebraic operations, this condition yields the following expression:
\begin{equation}
\frac{\lambda_{1}{\sum_{k,l}{P(k,l)k^{2}\frac{l^{2}\sigma_{2}^{2}\beta_{2}^{a}\beta_{1}^{a}\beta_{1}^{b}+l\sigma_{2}(\eta_{2}\mu_{2}\beta_{1}^{a}+\beta_{1}^{b}(\beta_{1}^{a}\mu_{1}+\beta_{2}^{a}\mu_{2}))+\mu_{2}(\eta_{1}\mu_{1}+\eta_{2}\mu_{2})}{l^{2}\sigma_{2}^{2}\beta_{2}^{a}\eta_{1}+l\sigma_{2}(\eta_{1}\mu_{1}+\eta_{2}\mu_{2}+ \beta_{2}^{a}\eta_{1}\mu_{2})+\mu_{2}(\eta_{1}\mu_{1}+\eta_{2}\mu_{2})}}}}{\mu_{1}\langle k \rangle}>1
\label{condition}
\end{equation} 
that allows us to derive the epidemic threshold as:
%
\begin{equation}
\lambda_{1}^{c}(\sigma_{2})=\mu_{1}\frac{\langle k \rangle}{\sum_{k,l}{P(k,l)k^{2}\frac{l^{2}\sigma_{2}^{2}\beta_{2}^{a}\beta_{1}^{a}\beta_{1}^{b}+l\sigma_{2}(\eta_{2}\mu_{2}\beta_{1}^{a}+\beta_{1}^{b}(\beta_{1}^{a}\mu_{1}+\beta_{2}^{a}\mu_{2}))+\mu_{2}(\eta_{1}\mu_{1}+\eta_{2}\mu_{2})}{l^{2}\sigma_{2}^{2}\beta_{2}^{a}\eta_{1}+l\sigma_{2}(\eta_{1}\mu_{1}+\eta_{2}\mu_{2}+ \beta_{2}^{a}\eta_{1}\mu_{2})+\mu_{2}(\eta_{1}\mu_{1}+\eta_{2}\mu_{2})}}}
\label{SecundaryThreshold2}
\end{equation} 
%
Looking at the latter expression $-$which contains the underlying topologies in a more intricate way than for the uncoupled, classical case$-$, the threshold dependence on disease 2's prevalence via $\sigma_{2}$ becomes explicit. If we evaluate $\lambda_{1}^{c}(\sigma_{2}=0)$ we recover the classical result  $\lambda_{1}^{c}=\mu_{1}\langle k\rangle/\langle k^{2}\rangle$ \cite{vespiromu,vrpre}. Therefore, in the following we will refer to this baseline case as primary threshold, $\lambda_{1}^{c}(0)$, whereas the more general case will be referred to as the secondary threshold, $\lambda_{1}^{c}(\sigma_{2})$ (with $\sigma_{2}>0$). Obviously, the same stands for the primary ($\lambda_{2}^{c}(0)$) and secondary thresholds ($\lambda_{2}^{c}(\sigma_{1})$) of the second disease.

A particular case for the topologies on top of which both diseases are spreading corresponds to the homogeneous mean field version of the system, i.e.,  $P(k,l)=\delta(k-k_{o})\delta(l-l_{o})$, for which the last expression can be rewritten as:
\begin{equation}
{\color{black}\lambda_{1}^{c}}=\frac{\lambda_{2}l_{o}}{k_{o}}\frac{\eta_{1}\mu_{1}(\beta_{2}^{a}(\lambda_{2}l_{o}-\mu_{2})+\mu_{1})+\eta_{2}\mu_{2}\mu_{1}}{\mu_{2}(\eta_{2}\mu_{2}+\eta_{1}\mu_{1})+(\lambda_{2}l_{o}-\mu_{2})(\beta_{2}^{a}\beta_{1}^{a}\beta_{1}^{b}(\lambda_{2}l_{o}-\mu_{2})+\beta_{1}^{a}\beta_{1}^{b}\mu_{1}+\eta_{2}\mu_{2}\beta_{1}^{a}+\beta_{2}^{a}\beta_{1}^{b}\mu_{2})}
\label{SecondaryThreshold1}
\end{equation}
An independent derivation of this expression can be obtained by analyzing the Jacobian matrix of the homogeneous mean field system analogous to Eq. (\ref{sistema}) as it is shown in the Appendix A.

\subsubsection{Phase diagrams}

In order to explore the quality of the threshold prediction of our model, we have designed a Monte-Carlo simulation scheme in which a single state transition is allowed per individual per time step. First, infected individuals will eventually spread the disease(s) that they carry. As double events are not allowed at a single time step, forbidden double transitions from {\color{black}$SS$} to $II$ are resolved by choosing the disease that an individual will catch proportionally to the status of her infected neighbors: the more neighbors one individual has, say, carrying disease 1, the more likely she will become infected with disease 1 rather than with disease 2. After the spreading loop is completed for both diseases, infected nodes who have not suffered any contagion at the present time step eventually get back to the susceptible state of the disease(s) they carry. To avoid forbidden double transitions from the $II$ class to the {\color{black}$SS$} class, in those cases the only disease the individual is going to recover from is also chosen stochastically, according to the probabilities $p_{1}=\eta_{1}\mu_{1}/(\eta_{1}\mu_{1}+\eta_{2}\mu_{2})$ for the first disease, and $p_{2}=1-p_{1}$ for the second one. 

Let us then explore first a simple scenario in which we assume that the dynamical effects of one disease on the other are totally symmetric. In terms of the parameters of the model, this implies that $\beta_{1}^{a}=\beta_{2}^{a}=\beta_{1}^{b}=\beta_{2}^{b}\equiv\beta$ and $\eta_{1}=\eta_{2}\equiv\eta$. In this case, let's focus on two opposite scenarios: i) mutual enhancement: $\eta<1$ and $\beta>1$ and ii) {\color{black}mutual impairment} $\eta>1$ and $\beta<1$. In the case of mutual enhancement, individuals who are infected with the second disease spread and become infected with disease 1 more easily than those who are not (this is because $\beta>1$). In addition, since also $\eta<1$, infected individuals of disease 1 remain so for longer times if they are also infected with the second disease. These two effects imply that the appearance of disease 2 in the system enhances the spreading capabilities of disease 1. The reciprocal situation is also true, as the interaction between both diseases is symmetric. Finally, in the case of {\color{black}mutual impairment}, the effects on the infectiousness and recovery rates are the opposite, and so the appearance of one of the diseases at a certain prevalence impairs the spreading of the other disease. In Figures \ref{fig2}--\ref{fig3}, we represent the prevalence of each disease, as a function of the baseline infectiousnesses $(\lambda_{1},\lambda_{2})$ for a given set of parameters after the introduction of infection seeds in the order shown. The networks through which diseases spread are, for this first case, two uncorrelated Erd\"os-Renyi graphs.

For the case of mutual enhancement {\color{black}(Fig. \ref{fig2})}, given our set of parameters, the analytically-obtained curves for the secondary threshold remain below the primary thresholds, leading to the appearance of two regions in the plane $(\lambda_{1},\lambda_{2})$, for which it is verified that $\lambda_{1}^{c}(\sigma_{2})<\lambda_{1}<\lambda_{1}^{c}(0)$ and $\lambda_{2}^{c}(\sigma_{1})<\lambda_{2}<\lambda_{2}^{c}(0)$, respectively. The dynamical relevance of these regions is that within them, the appearance of an outbreak of one of the diseases is conditional to the previous installation of the other infection in the system. In this way we can observe that, after an initial seed of $IS$ individuals, disease 1 does not become endemic in the region $\lambda_{1}^{c}(\sigma_{2})<\lambda_{1}<\lambda_{1}^{c}(0)$ (fig. \ref{fig2}, panel a), but then, after the outbreak of the second disease in the network, the same seed leads disease 1 to become endemic in that same region (fig. \ref{fig2} panel C) as predicted by our model. Regarding the conjugate region in which $\lambda_{2}^{c}(\sigma_{1})<\lambda_{2}<\lambda_{2}^{c}(0)$, we can see in figure \ref{fig2}, panel $e$, that disease 2 directly becomes endemic after the introduction of an infection seed $SI$ due to the fact that, previously, disease 1 was already introduced in the system. 

The situation for the {\color{black}mutual impairment} case is the opposite, and the secondary thresholds remain, in this case, above the primary ones. So, in this scenario, we have another couple of relevant regions in which $\lambda_{1}^{c}(0)<\lambda_{1}<\lambda_{1}^{c}(\sigma_{2})$ and $\lambda_{2}^{c}(0)<\lambda_{2}<\lambda_{2}^{c}(\sigma_{1})$, respectively. In fig.\ \ref{fig3}, it is represented the behavior of the system under these conditions. In panel a, we can see how disease 1 becomes endemic after the introduction of an initial seed above its primary threshold. Then, after introducing a seed of disease 2, as shown in panel b for the area comprised between $\lambda_{1}^{c}(0)$ and $\lambda_{1}^{c}(\sigma_{2})$, the prevalence of disease 1 vanishes. In other words, in that region, the introduction of disease 2 makes it possible for the system to recover from disease 1. If we look at the behavior of the second disease in the region in which $\lambda_{2}^{c}(0)<\lambda_{2}<\lambda_{2}^{c}(\sigma_{1})$, we see how the disease is unable to become endemic as a consequence of the fact that the first disease has already been introduced in the system. This situation suggests that, as it has already been addressed in the context of computational sciences \cite{fire-fire}, the introduction of an infectious agent designed to immunize its host with respect to another, more harmful infection, might be a conceptually feasible option to reduce the prevalence of the latter, or even to eradicate it. This has also been recently reported in the context of multi-strain diseases, in which more than one strain of the same disease compete for the host population \cite{poletto,sahneh2}.

\begin{figure}
  \begin{center}
  \includegraphics[]{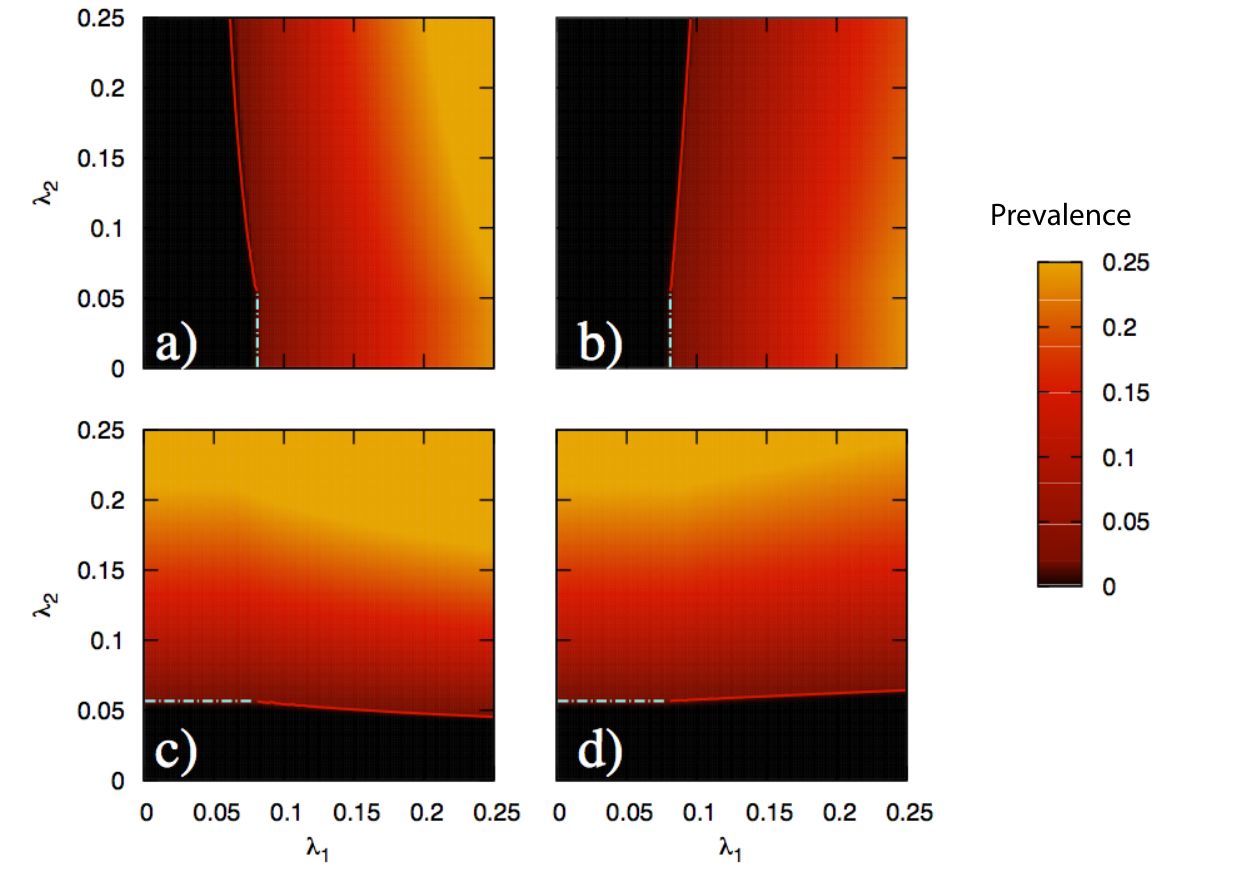}
  \end{center}
\caption{(color online) Panels a and c: reciprocally enhanced spreading with $\beta_{1}^{a}=\beta_{1}^{b}=\beta_{2}^{a}=\beta_{2}^{b}=1.3$, $\eta_{1}=\eta_{2}=0.8$, represented at the final stationary state. Panels b and d: reciprocally impaired spreading $\beta_{1}^{a}=\beta_{1}^{b}=\beta_{2}^{a}=\beta_{2}^{b}=0.8$, $\eta_{1}=\eta_{2}=1.3$, also at the final state. Scale-free networks are generated using the uncorrelated configuration model with $N_{1}=N_{2}=5000$ agents, $\langle k\rangle=4.00$,  and $\langle l\rangle=5.11$. The figure represents the final prevalence of diseases 1 (panels a and b) and 2 (panels c and d) in each case.}
\label{figure4}
\end{figure}

Once the dynamics of the model has been exhaustively characterized when the diseases spread over homogeneous networks, we move on and explore the influence of degree heterogeneity on the dynamics. To this end we also perform intensive numerical simulations in an analogous way, but using scale free graphs of the same size as before with different exponents. In Figure \ref{figure4} we represent the final prevalence for each disease in two configurations: reciprocally enhanced diseases $-$panels a and c$-$ and impaired spreading $-$panels b and d$-$.  In both cases, network 1 ($\gamma=2.7$, panels $a$ and $b$) has a greater power law exponent than network 2 ($\gamma=2.5$, panels $c$ and $d$). As it can be seen, for both diseases and for both configurations, secondary thresholds are closer to primary ones than in the case of homogeneous networks. 

\subsubsection{System sizes and epidemic thresholds: general case}

In the previous sections we have described the baseline cases in which none of the dynamical parameters vanishes, and both networks are, in each case, of the same kind --ER or uncorrelated SF graphs--. A relevant theoretical question yet remains unanswered, i.e., how the epidemic thresholds behave when the system size grows and eventually reaches the thermodynamic limit.  Regarding this question, we present an exhaustive analysis in Appendix B, in which it is shown what are the conditions that lead to have vanishing small secondary thresholds as a function of the underlying topologies and some of the dynamical parameters $-$ note, however, that this question is of interest from a theoretical viewpoint, as strictly speaking all real systems are finite and thus an effective epidemic threshold exits. {\color{black} In addition, our approach is based on an HMF approximation, and that means that influences of dynamical correlations on the analysis are neglected}. 

The results show that the secondary epidemic threshold associated to any of the diseases that is spreading over a scale-free network with $2<\gamma \le 3$ vanishes at the thermodynamic limit, regardless of the topology of the network on top of which its conjugate disease propagates and of the values of the dynamical parameters. In an analogously robust way, power laws with $\gamma > 3$ $-$or homogeneous degree distributions$-$ yield finite, non vanishing secondary thresholds at the thermodynamic limit, regardless of the conjugate topologies or parameter values, with some exceptions. 

The relevance of this result relies on the fact that the model predicts the same behavior for the epidemic thresholds in heterogeneous and homogeneous networks {\color{black}as compared with HMF models} of uncoupled (single) diseases that spread over simple networks. In addition, our analysis shows that the eventual vanishing of the epidemic thresholds for infinite systems is only determined by the topology of the network under consideration rather than by any possible coupling with another disease that spreads over any other possible conjugate network within our model framework. However, there is an exception to this general behavior which is meaningful from an epidemiological viewpoint. This is the case when both diseases spread over two highly and positively correlated scale-free networks with composed degree distribution $P(k,l)=\delta(k-l)\alpha k^{-\gamma}$, where $\delta$ stands for the Kronecker $\delta$-function and $\alpha$ is a normalization constant. In that situation, if we focus, for example, in disease 1, there exist two different interaction schemes of interest for which we recover finite epidemic thresholds $\lambda_{1}^{c}>0$ at the thermodynamic limit even for scale-free graphs with $2<\gamma\leq3$:

\begin{itemize}
\item Case 1: Individuals infected with disease 2 become immune to infection by disease 1. $\beta_{1}^{a}=0$ ; $(\beta_{2}^{a},\beta_{2}^{b},\beta_{2}^{b})$ are free parameters.
\item Case 2: If $\beta_{1}^{a}\neq0$, individuals infected with both diseases can not cause contagion of disease 1:  $\beta_{1}^{b}=0$. In addition, disease 1 can not cause total immunity to disease 2: $\beta_{2}^{a}\neq0$.  $\beta_{2}^{b}$ is a free parameter.
\end{itemize}

To illustrate this situation, we take as an example a particular case of the first scheme, a mutual cross-immunity scenario given by $\beta_{1}^{a}=\beta_{2}^{a}=0$. In order to point out the role of inter-layer degree correlations on this effect, we can directly compare the expression for the threshold when both networks are totally correlated with the analogous expression derived from {\color{black}an uncorrelated} combined degree distribution:
 
\begin{equation}
P(k,l)=\delta(k-l)\alpha k^{-\gamma}\rightarrow\lambda_{1}^{c}(\sigma_{2})=\mu_{1}
\frac
{\langle k \rangle}
{\sum_{k}{\alpha\frac{1}{k^{2-\gamma}}\frac{\mu_{2}}{k\sigma_{2}+\mu_{2}}}}
\label{reducedThreshold1}
\end{equation} 
\begin{equation}
P(k,l)=\alpha k^{-\gamma}l^{-\Gamma}\rightarrow\lambda_{1}^{c}(\sigma_{2})=\mu_{1}
\frac
{\langle k \rangle}
{\sum_{k,l}{\alpha\frac{1}{k^{2-\gamma}}\frac{1}{l^{\Gamma}}\frac{\mu_{2}}{l\sigma_{2}+\mu_{2}}}}
\label{reducedThreshold2}
\end{equation} 

\begin{figure}
  \begin{center}
  \includegraphics[width=\columnwidth]{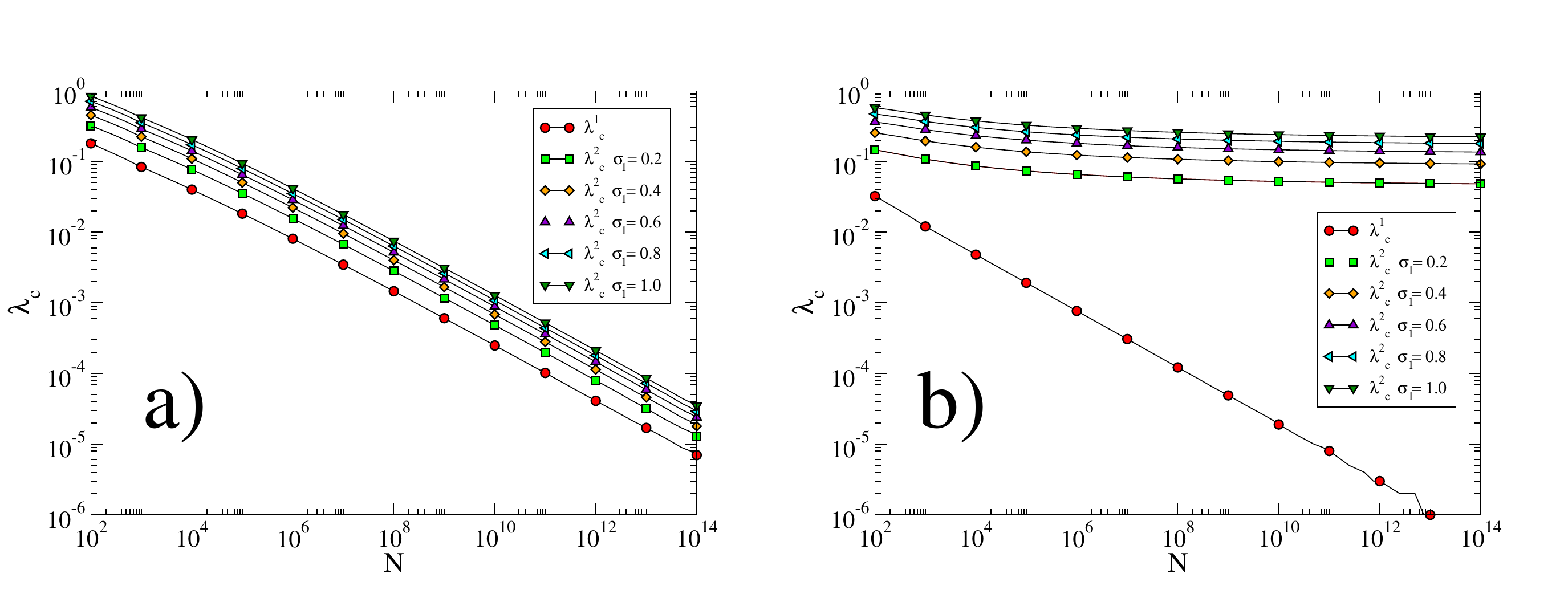}
  \end{center}
\caption{(color online) Primary and Secondary threshold values as a function of the size of the systems $N$ for different values of $\Theta$: $0.0$ (primary threshold), $0.2$, $0.4$, $0.6$, $0.8$ and $1.0$. Networks are uncorrelated scale free graphs (panel a) without correlations between them --panel a-- or with them --panel b, fully correlated--. Diseases interact according to a full cross-immunity scheme $\beta_a^{1}=\beta_{a}^{2} = 0$ and $\gamma=\Gamma=2.5$. The rest of the parameters are the same as in figure \ref{fig2}.}
\label{figure5}
\end{figure}

In Figure \ref{figure5}, we represent the values predicted by these expressions for different network sizes, when $\gamma=\Gamma=2.5$. As we can see in panel a, for uncorrelated networks, regardless of the value of $\sigma_{2}$, the threshold continuously decreases as we increase network sizes. The result thus shows that the existence of a coupling with another disease present in the system with a certain prevalence proportional to $\sigma_{2}$, does not play any role, since the degree heterogeneities are still the main reason leading to the vanishing of the threshold at the thermodynamic limit. This picture turns out to be remarkably different when we introduce positive, strong correlations between the two networks. In that case, as we can see in panel b, the appearance of the second disease, characterized by a certain prevalence level $\sigma_{2}>0$, implies a sudden change in the behavior of the threshold, that does not vanish anymore, even when $N\rightarrow \infty$.


\begin{figure}
  \begin{center}
  \includegraphics[]{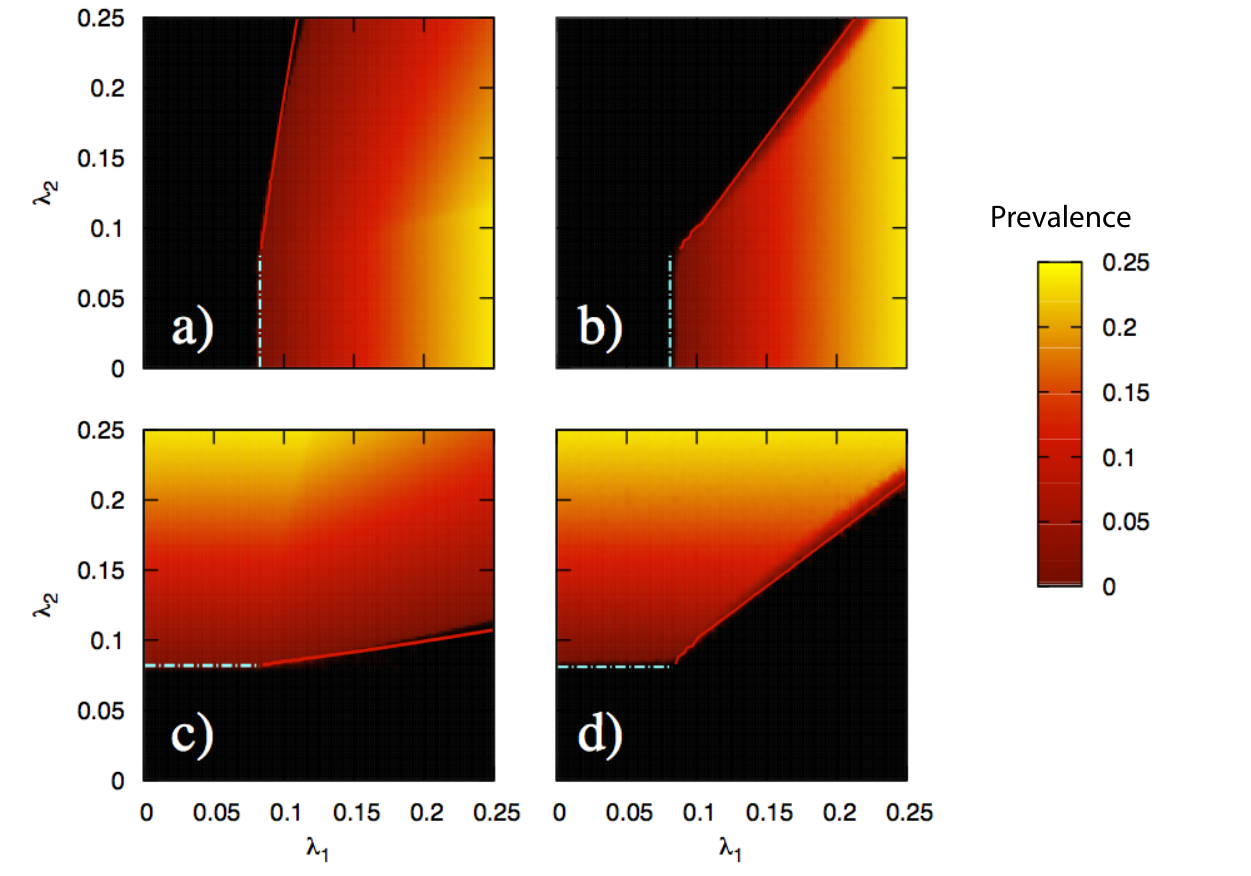}
  \end{center}
\caption{(color online) Effect of degree correlations between the two scale-free networks on the steady prevalence levels. Dynamical parameters: $\mu_{1}=\mu_{2}=0.75$, $\beta_{1}^{a}=\beta_{2}^{a}=0$. The rest of parameters are irrelevant, as no $II$ individual will appear in the system. Panels a and c: final prevalence of diseases 1 and 2, respectively for uncorrelated scale-free networks: $P(k,l)=P(k)P(l)=\alpha k^{-\gamma}l^{-\gamma}$. Panels b and d: final prevalence of diseases 1 and 2, respectively for fully correlated scale-free networks: $P(k,l)=\alpha \delta(k-l)k^{-\gamma}$. In both cases $\gamma=2.5$.}
\label{figure6}
\end{figure}

The influence of degree correlations between networks for this case of full cross-immunity becomes evident also at finite sizes, since the differences between primary and secondary thresholds, as seen in Fig. \ref{figure6}, are also greatly amplified. Another eventually relevant effect that can be observed in the last figure is that the transition that takes place at the epidemic threshold is much sharper in the case of correlated networks. All the previous results point out that the worst scenario for the spreading of a disease when it interacts with a second one that confers immunity to the former corresponds to the case in which there is a correlation between heterogeneous networks of contacts. This finding is essentially equivalent to what was found previously in \cite{fj10}, what we have shown here is that the effect comes to revert the vanishing threshold at the thermodynamic limit for a disease spreading on top of scale-free networks. In addition, we have found that it is not needed for this effect to take place that the second disease confers full immunity to disease one. 

{\color{black}Remarkably all these ``pathological'' cases can be identified without abandoning HMF descriptions. Beyond HMF, the behavior of epidemic thresholds vary with respect to mean field predictions as a consequence of dynamical correlations \cite{caste,eames}. Here, however, we identify that the interactions between diseases can modify the size scaling of thresholds from the classical mean field picture without the need to recurring to dynamical correlations, which, for certain cases, remarkably modify the whole picture at the thermodynamic limit, even in annealed networks in which adjacency matrices are only fixed on average, and so, dynamical correlations do not exist}.

\subsection{The SIR scenario}

In the previous sections we have studied the behavior of systems of interacting diseases that spread according to a SIS scheme, thus leading, above the epidemic threshold, to stationary, endemic states with a prevalence greater than zero. In the following, we will explore the case of transient, interacting epidemic phenomena.

In order to do so, we can extend the framework and describe the dynamics of two SIR epidemics interacting among them. In classical, non-interacting systems --either homogeneous or heterogeneous--, the resemblance of both types of models translates into a strong mathematical symmetry between them that yields identical expressions for the epidemic thresholds {\color{black}under mean field descriptions (i.e. when neglecting the effects of dynamical correlations)}. In this section we will see the way in which part of this symmetry is broken as a consequence of the interacting nature of the epidemic processes.
\subsubsection{Mathematical description}
 
\begin{figure}
\begin{center}
 \includegraphics[width=0.5\columnwidth]{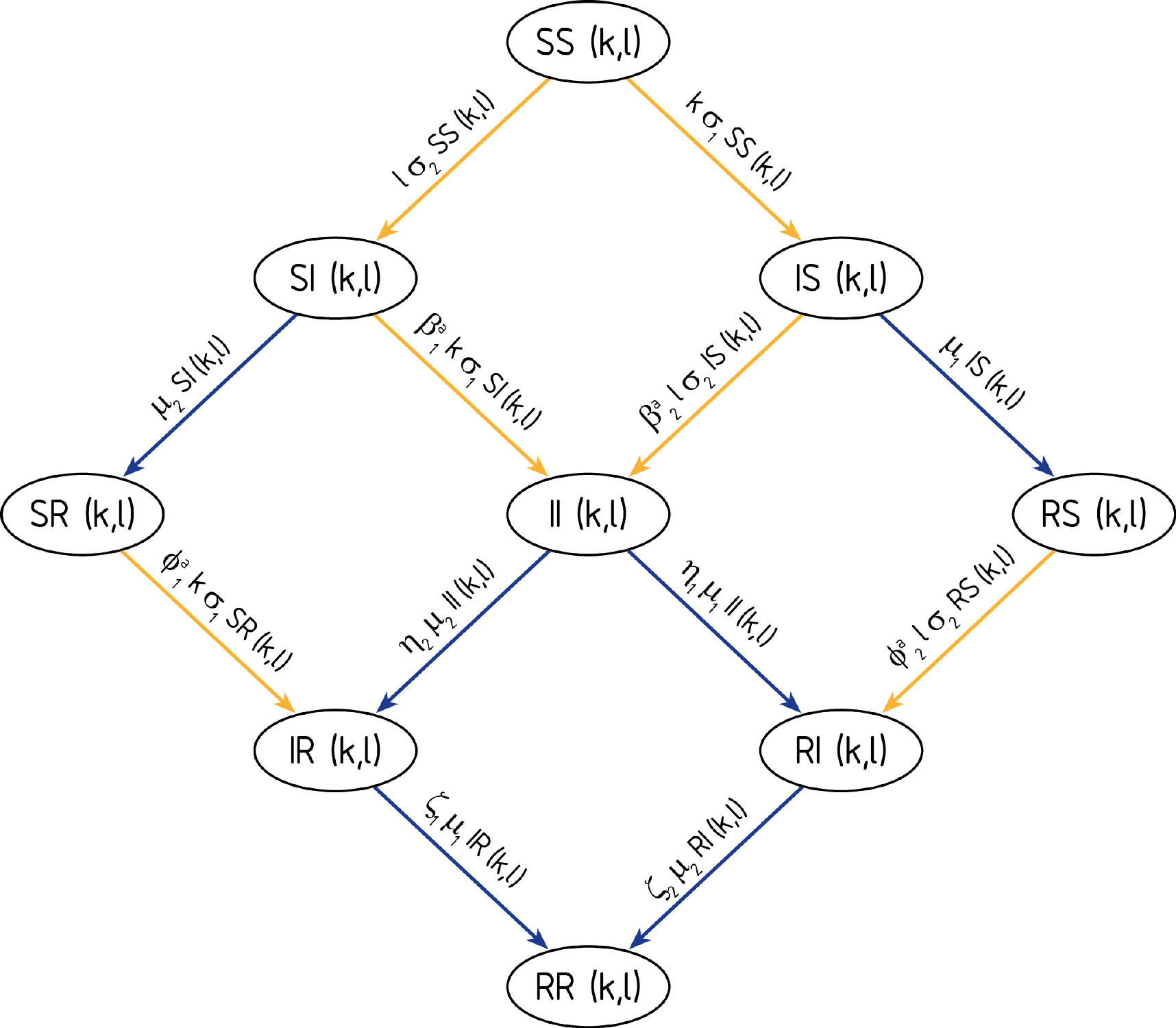}
\caption{(color online)Set of transitions allowed in the double SIR-SIR model. Yellow: contagion processes. Blue: recovery processes}
\label{sir-sir-scheme}
\end{center}
\end{figure}

In this case, the first new aspect to note is that not just the condition of being infected with one disease can modify subjects' susceptibility to the conjugate infection, but also whether they have been infected and recovered. For instance, this might represent situations in which some kind of immunological memory is acquired after the first infection $-$for example, partial immunity in front of other strains is gained after suffering an influenza infection, specially if both are phylogenetically close enough \cite{tpl13}. This new phenomenology translates into the need of introducing new parameters (see Table \ref{new-parameters-table}) describing new eventual interactions and transitions, as shown in Fig \ref{sir-sir-scheme}. The system of equations describing the dynamics is now:

\begin{table}
\begin{center}
\begin{tabular}{c|p{87mm}l}
\hline \hline
New parameter & Dynamical meaning \\\hline
$\phi_{1}^{a}$ & Variation of disease $1$ infectiousness due to the fact that the susceptible individual exposed to disease $1$ has been infected and recovered from disease $2$\\\hline
$\phi_{2}^{a}$ & Variation of disease $2$ infectiousness due to the fact that the susceptible individual exposed to disease $2$ has been infected and recovered from disease $1$\\\hline
$\phi_{1}^{b}$ & Variation of disease $1$ infectiousness due to the fact that the spreader has been infected and recovered from disease $2$\\\hline
$\phi_{2}^{b}$ & Variation of disease $2$ infectiousness due to the fact that the spreader has been infected and recovered from disease $1$\\\hline
$\zeta_{1}$ & Variation of disease $1$ recovery rate for individuals that have been infected and recovered from disease $2$\\\hline
$\zeta_{2}$ & Variation of disease $2$ recovery rate for individuals that have been infected and recovered from disease $1$\\\hline\hline
\end{tabular}
\caption{Parameters describing the influence of R classes on the conjugate infection}
\label{new-parameters-table}
\end{center}
\end{table}

\begin{center}
\begin{eqnarray}
\dot{SS}(k,l)=-(k\sigma_{1}+l\sigma_{2})SS(k,l)\\
\dot{IS}(k,l)=k\sigma_{1}SS(k,l)-\beta_{2}^{a}l\sigma_{2}IS(k,l)-\mu_{1}IS(k,l)\\
\dot{SI}(k,l)=l\sigma_{2}SS(k,l)-\beta_{1}^{a}k\sigma_{1}SI(k,l)-\mu_{2}SI(k,l)\\
\dot{II}(k,l)=\beta_{1}^{a}k\sigma_{1}SI(k,l)+\beta_{2}^{a}l\sigma_{2}IS(k,l)-(\eta_{1}\mu_{1}+\eta_{2}\mu_{2})II(k,l)\\
\dot{RS}(k,l)=\mu_{1}IS(k,l)-\phi_{2}^{a}l\sigma_{2}RS(k,l)\\
\dot{SR}(k,l)=\mu_{2}SI(k,l)-\phi_{1}^{a}k\sigma_{1}SR(k,l)\\
\dot{RI}(k,l)={\color{black}\phi_{2}^{a}l\sigma_{2}RS(k,l)}+\eta_{1}\mu_{1}II(k,l)-\zeta_{2}\mu_{2}RI(k,l)\\
\dot{IR}(k,l)={\color{black}\phi_{1}^{a}k\sigma_{1}SR(k,l)}+\eta_{2}\mu_{2}II(k,l)-\zeta_{1}\mu_{1}IR(k,l)
\label{sir-system}
\end{eqnarray}
\end{center}

where all the parameters and variables that were present in the SIS formulation retain their original meaning, with the nuance that now $\sigma_{1}$ and $\sigma_{2}$ have an extra term related to the appearance of classes $IR$ and $RI$:
\begin{center}
\begin{eqnarray}
\sigma_{1}=\lambda_{1}(\theta_{1}^{IS}+\beta_{1}^{b}\theta_{1}^{II}+\phi_{1}^{b}\theta_{1}^{IR})\\\nonumber
\sigma_{2}=\lambda_{2}(\theta_{2}^{SI}+\beta_{2}^{b}\theta_{2}^{II}+\phi_{2}^{b}\theta_{2}^{RI})
\label{sigmas-new}
\end{eqnarray}
\end{center} 
where $\theta_{1}^{IR}$ is the probability of a link of network 1 pointing to an $IR$ node and $\theta_{2}^{RI}$ is the probability of a link of network 2 pointing to a node in the $RI$ state.

\subsubsection{Epidemic thresholds}
 
In order to characterize the epidemic threshold of disease 2, we need to study the dynamics of the system around any point in which no infected individual of disease 2 has yet been introduced and so $(SI{(k,l)},II{(k,l)},RI{(k,l)})=(0,0,0)\forall (k,l)$ as well as $\theta_{2}^{SI}=\theta_{2}^{II}=\theta_{2}^{RI}=0$. Around such a disease free point -either fixed or not-, it is trivial to see that all the partial derivatives of classes $SI({k,l}),II({k,l})$ and $RI({k,l})$ with respect to the rest of dynamic classes vanish. This makes the subset of variables $\{SI({k,l}),II({k,l}),RI({k,l})\}$ locally autonomous around that point, and so, its linearization might serve us to address the stability inversion yielding the emergence of the epidemic threshold.

In addition, the dimensionality of the system can be greatly reduced, if we write the equations driving the time evolution of the probabilities $\theta_2^{SI}$, $\theta_2^{II}$, and $\theta_2^{RI}$, as follows:
 
{\color{black} 
 \begin{center}
\begin{eqnarray}
\dot{\theta_{2}^{SI}}=\frac{\sum_{k,l}{P(k,l)l \dot{SI}}}{\langle l\rangle}=\frac{\langle l^{2}SS\rangle}{\langle l\rangle}\lambda_{1}(\theta_{2}^{SI}+\beta_{2}^{b}\theta_{2}^{II}+\phi_{2}^{b}\theta_{2}^{RI})-\beta_{1}^{a}\frac{\langle kl\rangle}{\langle l\rangle}\theta_{2}^{SI}\lambda_{1}(\theta_{1}^{IS}+\beta_{1}^{b}\theta_{1}^{II}+\phi_{1}^{b}\theta_{1}^{IR})-\mu_{2}\theta_{2}^{SI}\\\nonumber
\dot{\theta_{2}^{II}}=\frac{\sum_{k,l}{P(k,l)l \dot{II}}}{\langle l\rangle}=\frac{\langle kl\rangle}{\langle l\rangle}\theta_{2}^{SI}\beta_{1}^{a}\lambda_{1}(\theta_{1}^{IS}+\beta_{1}^{b}\theta_{1}^{II}+\phi_{1}^{b}\theta_{1}^{IR})+\frac{\langle l^{2}IS\rangle}{\langle l\rangle} \beta_{2}^{a}\lambda_{2}(\theta_{2}^{SI}+\beta_{2}^{b}\theta_{2}^{II}+\phi_{2}^{b}\theta_{2}^{RI})-(\eta_{1}\mu_{1}+\eta_{2}\mu_{2})\theta_{2}^{II}\\\nonumber
\dot{\theta_{2}^{RI}}=\frac{\sum_{k,l}{P(k,l)l \dot{RI}}}{\langle l\rangle}=\frac{\langle l^{2}RS\rangle}{\langle l\rangle} \phi_{2}^{a}\lambda_{2}(\theta_{2}^{SI}+\beta_{2}^{b}\theta_{2}^{II}+\phi_{2}^{b}\theta_{2}^{RI})+\eta_{1}\mu_{1}\theta_{2}^{II}-\zeta_{2}\mu_{2}\theta_{2}^{RI}
\label{thetas-system}
\end{eqnarray}
\end{center}
}
where we have substituted $\langle klSI\rangle$ by $\langle kl\rangle\theta_{2}^{SI}$, approximation which is valid around the point $(SI{(k,l)},II{(k,l)},RI{(k,l)})=(0,0,0)\forall (k,l)$. Obviously, as happened for $\{SI({k,l}),II({k,l}),RI({k,l})\}$, all of the partial derivatives of $\theta_2^{SI}$, $\theta_2^{II}$, and $\theta_2^{RI}$ with respect to variables unrelated to $\theta_2^{SI}$, $\theta_2^{II}$, and $\theta_2^{RI}$ vanish, which allows us to study the stability of the system with respect to disease 2 by linearizing the system: $(\dot{\theta_2^{SI}},\dot{\theta_2^{II}},\dot{\theta_2^{RI}})=f(\theta_2^{SI},\theta_2^{II},\theta_2^{RI})$. The corresponding Jacobian matrix $J$ can be calculated this way, and, by evaluating the condition $J=0$ for stability shift, the epidemic threshold takes its final value:

{\footnotesize
\begin{center}
\begin{eqnarray}
\nonumber
\lambda_{2}^{c}(\langle l^{2}SS\rangle,\langle l^{2}IS\rangle,\langle l^{2}RS\rangle,\sigma_{1})=\\ 
\frac {(\eta_{2}\mu_{2}+\eta_{1}\mu_{1})(\mu_{2}\langle l\rangle+\beta_{1}^{a}\sigma_{1}\langle kl\rangle)\zeta_{2}\mu_{2}\langle l\rangle}
{(\eta_{2}\mu_{2}+\eta_{1}\mu_{1})
(\zeta_{2}\mu_{2}\langle l\rangle\langle l^{2}SS\rangle
+\phi_{2}^{a}\phi_{2}^{b}\langle l^{2}RS\rangle(\mu_{2}\langle l\rangle+\beta_{1}^{a}\langle kl\rangle\sigma_{1}))
+ (\beta_{2}^{b}\zeta_{2}\mu_{2}+\phi_{2}^{b}\eta_{1}\mu_{1})
(\beta_{1}^{a}\langle kl\rangle\sigma_{1}\langle l^{2}SS\rangle+
\beta_{2}^{a}(\mu_{2}\langle l\rangle+\beta_{1}^{a}\langle kl\rangle\sigma_{1})\langle l^{2}IS\rangle)}
\label{sir-secundary}
\end{eqnarray}
\end{center}}
which depends on the initial prevalence of disease 1 via $\sigma_{1}$, $\langle l^{2}IS\rangle$, $\langle l^{2}RS\rangle$ (and $\langle l^{2}SS\rangle$). In the case of having non concurrent outbreaks, we have that the second disease arrives to the system after the outbreak of the first disease has come to an end. In that case, we have that $IS({k,l})=0$ and $RS({k,l})=R_{\infty,k,l}^{1}$ $\forall (k,l)$, where $R_{\infty,k,l}^{1}$ is the fraction of recovered individuals at the end of an outbreak of disease one alone, in the composed degree class $(k,l)$. In such a case, the problem is much simpler, as $\sigma_{1}=\langle l^{2}IS\rangle=0$, and {\color{black}$\langle l^{2}SS\rangle=\langle l^{2}\rangle-\langle l^{2}R_{\infty,k,l}^{1}\rangle$}; and the threshold reads as:

{\color{black}
\begin{equation}
\lambda_{2}^{c}(\langle l^{2}\rangle,\langle l^{2}R_{\infty,k,l}^{1}\rangle)=\frac{\mu_{2}\langle l\rangle}{\langle l^{2}\rangle+\frac{(\phi_{2}^{a}\phi_{2}^{b}-\zeta_{2})}{\zeta_{2}}\langle l^{2}R_{\infty,k,l}^{1}\rangle}.
\label{sir-secundary-non-concurrent}
\end{equation}
}
Obviously, if disease 1 has not yet appeared in the system, the threshold for disease 2 becomes {\color{black}$\lambda_{2}^{c}=\frac{\mu_{2}\langle l\rangle}{\langle l^{2}\rangle}$}, as in the classical, non interacting {\color{black}HMF} case \cite{mpsv02}. As done for the SIS model, we refer to the latter expression as the primary threshold of disease 2, in counterposition to the secondary threshold of Eq. (\ref{sir-secundary}).

\begin{figure}
\begin{center}
\includegraphics[width=0.5\columnwidth]{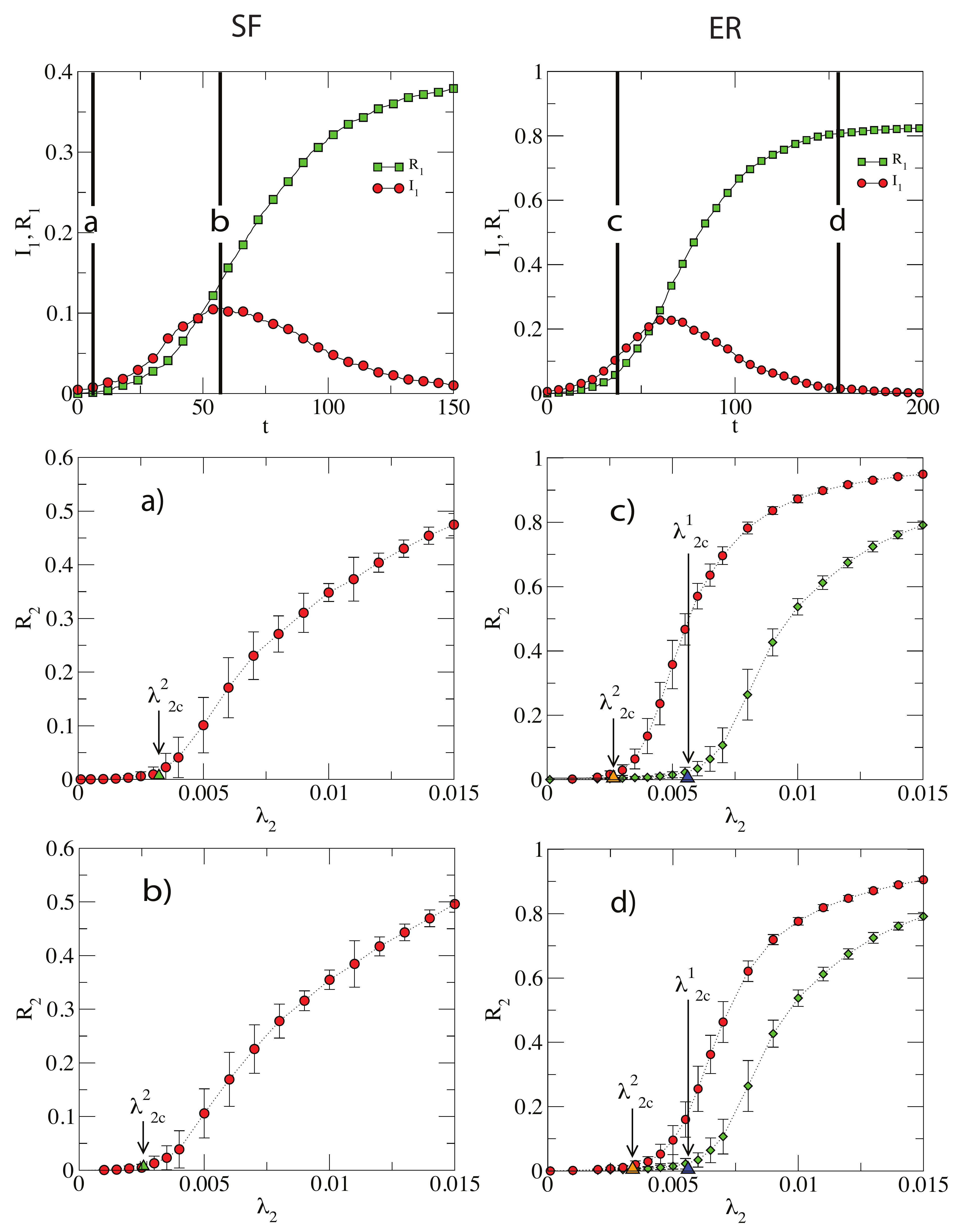}
\caption{(color online) Epidemic thresholds in the SIR-SIR interacting epidemic model. Upper panels: temporal evolution of disease 1, in a scale free (SF, left) and in an Erdos-Renyi network (ER, right): $I_{1}=\sum_{k,l}{P(k,l)(IS_{k,l}+II_{k,l}+IR_{k,l})(t)}$, $R_{1}=\sum_{k,l}{P(k,l)(RS_{k,l}+RI_{k,l}+RR_{k,l})(t)}$. Disease one enhances the spreading of disease two in both cases, but still is not influenced by the second infection. Lower panels: total number of affected individuals by disease 2 as a function of $\lambda_{2}$ in each network: $R_{2}^{\infty}=\lim_{t\rightarrow\infty}{\sum_{k,l}{P(k,l)(SR_{k,l}+IR_{k,l}+RR_{k,l})(t)}}$. Dynamical parameters: panels a, b: scale free network: $N=5000$, $\langle k\rangle=4.00$,$\langle l\rangle=5.11$, $\lambda_{1}=0.02$ $\mu_{1}=\mu_{2}=0.05$, $\beta_{2}^{a}=\beta_{2}^{b}=\phi_{2}^{a}=\phi_{2}^{b}=1.3$ $\eta_{2}=\zeta_{2}=0.8$ ,  panels c,d: Erdos-Renyi network: $N=5000$, $\langle k\rangle=7$,$\langle l\rangle=8$, $\lambda_{1}=0.02$ $\mu_{1}=\mu_{2}=0.05$, $\beta_{2}^{a}=\beta_{2}^{b}=3$, $\phi_{2}^{a}=\phi_{2}^{b}=1.2$, $\eta_{2}=0.33$, $\zeta_{2}=0.8$. As said before, in both cases, disease 1 is not influenced by disease 2, and so: $\beta_{1}^{a}=\beta_{1}^{b}=\phi_{1}^{a}=\phi_{1}^{a}=\eta_{1}=\zeta_{1}=1$.}
\label{lambda-2-c}
\end{center}
\end{figure}

A remarkable property of the threshold for the SIR scenario is that its dependence on the dynamical state of the conjugate disease is more complex than in the SIS case. Specifically, once an outbreak of one influencing infection is unfolding, the epidemic thresholds of the other disease may vary with time in non trivial ways, depending on the nature and intensity of the different interaction mechanisms present. Figure \ref{lambda-2-c} shows results from numerical simulations illustrating the previous phenomenology and the agreement with the analytical thresholds. Specifically, each panel represents the case in which the infection seed of the second disease is introduced in different moments for each topology, coinciding with different stages of the outbreak of disease 1: early phases (panel a, SF and panel c, ER), outbreak's peak (panel b, SF) and at the end of the outbreak (panel d, ER). As we can see, regardless of the topology, the parameters' values or the moment of appearance of the infection seed, our model adequately foresees the epidemic threshold and its variations with time. As in the SIS case, the influence of the interactions on the epidemic threshold is smaller in the case of scale free networks, due to the smaller values of the primary thresholds in that case (in fact, primary thresholds have not been represented in panels a and b for the sake of clarity, because its values $\lambda_{2c}^{1} = 0.00316$ virtually overlaps the secondary ones). 

\section{Conditions for disease enhancement and impairment}

Finally, it is interesting to analyze what combinations of parameters give raise to situations in which there is either enhancement or impairment of the diseases, as well as some other cases in which both effects are possible. This can be done for both the SIS and the SIR scenarios by studying the sign of the difference between the primary and the secondary thresholds.

For the SIS case, if we focus, for example, on the second disease, we have:

\begin{equation}
\lambda_{2}^{c}(\sigma_{1})-\lambda_{2}^{c}(0)=\sum_{k,l}{C(k,l)\sigma_{1}
\left[
\beta_{1}^{a}(\eta_{2}-\beta_{2}^{a}\beta_{2}^{b})k\sigma_{1}
+ (\eta_{2}-\beta_{2}^{a}\beta_{2}^{b})\mu_{2}+\left(\eta_{1}(1-\beta_{2}^{a})+\beta_{1}^{a}(\eta_{2}-\beta_{2}^{b})  \right)
\right]}
\end{equation}
where $C(k,l)$ is always positive. In this sense, $\lambda_{2}^{c}(\sigma_{1})-\lambda_{2}^{c}(0)>0$ implies that disease 1 is reducing the epidemic threshold of disease 2, thus enhancing its spreading. In the opposite case $\lambda_{2}^{c}(\sigma_{1})-\lambda_{2}^{c}(0)<0$, disease 1 makes the secondary threshold for disease 2 to be bigger than the primary one, hence impairing its spreading. As we can see, the condition yielding one or another case involves a complex combination of the parameters. However, it is trivial to show that, provided that $\beta_{2}^{a}>1$, $\beta_{2}^{b}>1$ and $\eta_{2}<1$, disease 1 enhances disease 2 spreading for any value of $\sigma_{1}$ greater than zero. We call this scenario coherent enhancement, because all the interaction mechanisms contribute to enhance the spreading of disease 2. In a similar way, if $\beta_{2}^{a}<1$, $\beta_{2}^{b}<1$ and $\eta_{2}>1$, the interaction has the opposite sign regardless of $\sigma_{1}$, and we have a situation of coherent impairment. Noticeably, if none of these conditions is fulfilled, there exists the possibility that the sign of the influence that disease 1 exerts on the spreading of disease 2 depends on its prevalence via $\sigma_{1}$ (see below). In Fig. \ref{deltas} we have represented $\lambda_{2}^{c}(\sigma_{1})-\lambda_{2}^{c}(0)$ for different parameter combinations that cover all possible phenomenologies.

\begin{figure}
\begin{center}
\includegraphics[width=0.5\columnwidth]{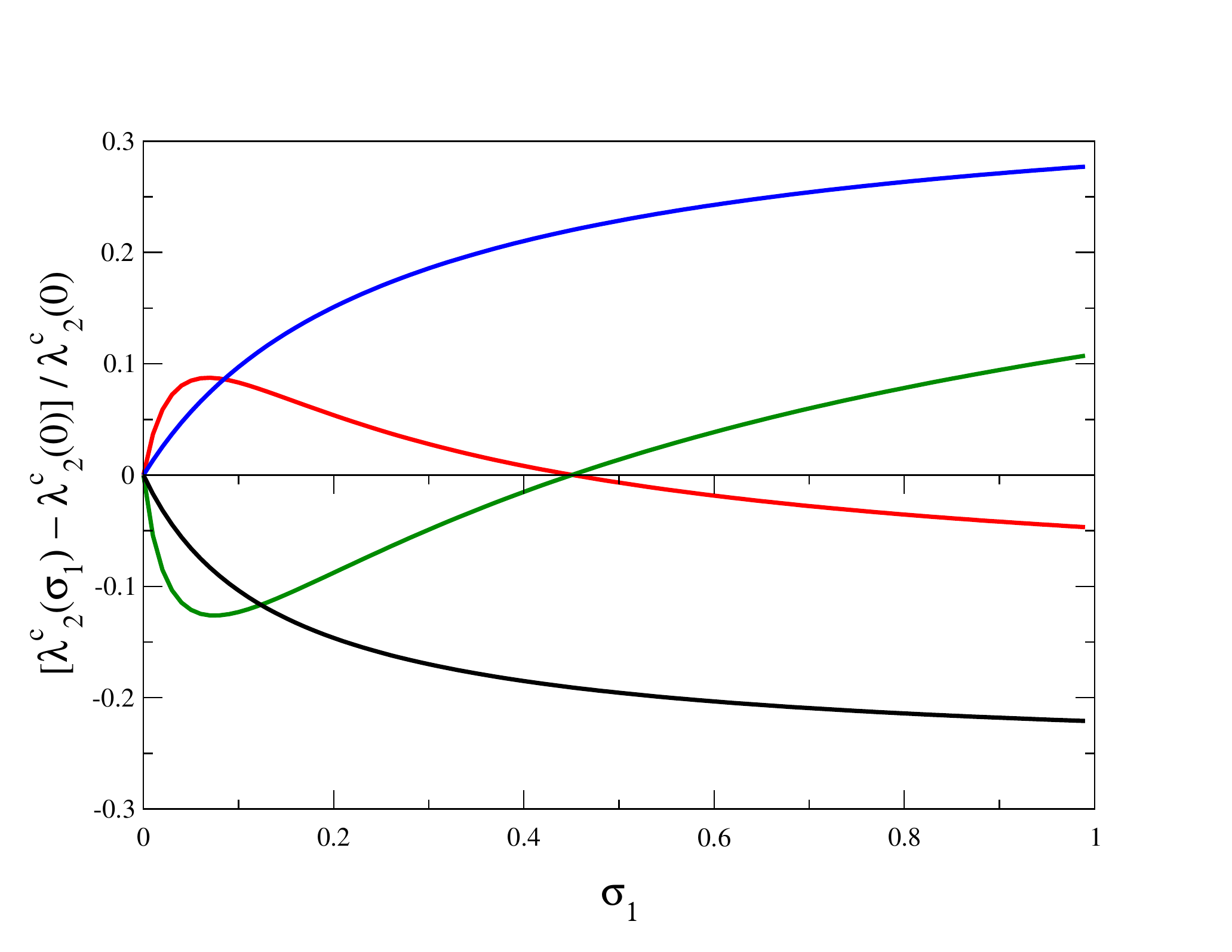}
\caption{Relative variation of epidemic threshold of disease 2 as a function of $\sigma_{1}$ for different parameter sets. Black: (coherent enhancement of disease 1 over disease 
       2): $\lambda_{1}=0.2$, $\mu_{1}=\mu_{2}=0.3$, $\beta_{2}^{a}=\beta_{2}^{b}=\beta_{1}^{a}=\beta_{1}^{b}=1.1$, $\eta_{1}=\eta_{2}=0.9$. 
Blue: (coherent impairment of disease 1 over disease 2): 
            $\lambda_{1}=0.2$, $\mu_{1}=\mu_{2}=0.3$, $\beta_{2}^{a}=\beta_{2}^{b}=\beta_{1}^{a}=\beta_{1}^{b}=0.9$, $\eta_{1}=\eta_{2}=1.1$. 
   Red: $\lambda_{1}=0.2$, $\mu_{1}=0.4$, $\mu_{2}=0.2$, $\beta_{2}^{a}=\beta_{2}^{b}=1.5$, $\beta_{1}^{a}=4$,  $\beta_{1}^{b}=0.5$, $\eta_{1}=0.4$, $\eta_{2}=2$. 
Green: $\lambda_{1}=0.2$, $\mu_{1}=0.4$, $\mu_{2}=0.2$, $\beta_{2}^{a}=0.5$, $\beta_{2}^{b}=1.5$, $\beta_{1}^{a}=4$, $\beta_{1}^{b}=0.5$, $\eta_{1}=0.4$, $\eta_{2}=1$. The networks are two Erd\"os-Renyi graphs: $N_{1}=N_{2}=5000$ agents, $\langle k\rangle=7$,  $\langle l\rangle=8$.}
\label{deltas}
\end{center}
\end{figure}

For the SIR case, the proliferation of dynamical classes and the possibility that other mechanisms (i.e., those including R individuals) carry the interaction between both disease makes the equivalent expression more complex, but still derivable as:

\begin{eqnarray}
\lambda_{2}^{c} (\langle l^{2}SS\rangle,\langle l^{2}IS\rangle,\langle l^{2}RS\rangle,\sigma_{1})-\lambda_{2}^{c} (\langle l^{2}\rangle,0,0,0) &=&\\ \nonumber
C'[
\mu_{2}\langle l\rangle(\eta_{1}\mu_{1}+\eta_{2}\mu_{2})
\left(
\zeta_{2}\beta_{1}^{a}\sigma_{1}\langle kl\rangle\langle l^{2}\rangle+\zeta_{2}\mu_{2}\langle l\rangle\langle l^{2}IS\rangle
+((\zeta_{2}-\phi_{2}^{a}\phi_{2}^{b})\mu_{2}\langle l\rangle -\beta_{1}^{a}\langle kl\rangle\sigma_{1})\langle l^{2}RS\rangle
\right) &-&\\ \nonumber
-\langle l \rangle\mu_{2} (\phi_{2}^{b}\eta_{1}\mu_{1}+\beta_{2}^{b}\zeta_{2}\mu_{2})\left(
\beta_{1}^{a}\sigma_{1}\langle kl\rangle\langle l^{2}SS\rangle+\beta_{2}^{a}(\mu_{2}\langle l \rangle+\sigma_{1}\langle kl\rangle)\langle l^{2}IS\rangle
\right)
]
\end{eqnarray}

where $C'$ is always positive too.

As we have already said, in the interacting SIR model, once an outbreak of one disease has started, the temporal dependence of the epidemic threshold of the second infection depends on the dynamic state of the system as well as on the parameter values that account for the mechanisms of interaction that are present in the system and their intensities. In such a case, if the interaction between the diseases is mediated by the class R, (i.e. being recovered of one disease is what makes subjects dynamics with respect to the other infection to be altered, and so $\beta$ and $\eta$ parameters are equal to 1), the density of individuals belonging to the R class is a monotonically increasing function of time. This trivially implies that, in general, the closer to its end is the epidemic process for one disease, the deeper its impact on the conjugate one. However, if it is the I class the one that carries the interaction, it is easy to see that that interaction will take place if and only if the outbreak of disease two happens within a time window that -at least partially- overlaps with that characterizing the outbreak of disease 1. In such a situation, the interaction between the two diseases is a transitory property of the system, and the shift in the epidemic threshold crucially depends on the temporal co-occurrence of the outbreaks of both diseases.

A similar time-dependent shift of the epidemic thresholds can also take place in the SIS model, for example for the thresholds of disease 2, if the difference $\lambda_{2}^{c}(\sigma_{1})-\lambda_{2}^{c}(0)$ changes its sign during an outbreak of disease 1, as a consequence of the evolution of $\sigma_{1}$. This is the scenario given by the green and red curves in Fig. \ref{deltas}. However, it is worth remarking that such combinations of parameters are somehow pathological, as they imply the appearance of diverse non-coherent mechanisms of interaction (i.e. one disease simultaneously enhances subjects susceptibility to disease but, at the same time, it reduces their infectiousness, etc.) that are epidemiologically less plausible. On the contrary, in the SIR framework, this transitoriness is an intrinsic property of the system when the infected class is the responsible of influencing the dynamics of the conjugate disease. This transitory nature of the interactions constitutes an essential difference between the double SIR model with respect to the SIS one, which is at the root of the remarkable difference between their corresponding thresholds and the richer phenomenology of two interacting SIR-like processes. 

\section{Conclusions}

{\color{black}
In this work, we propose a composed HMF model to describe the simultaneous spreading of two diseases over the same population but driven by independent mechanisms of transmission taking place on different networks of contacts. Within this framework, we have thoroughly studied extensions of the SIS and SIR models, where the parameters defining the infectious and recovery transitions of one disease depend on the state of each node with respect to the conjugate disease, establishing in this way a coupling between both diffusion processes. Our modeling approach presents different advantages with respect to previous models \cite{kn11,fj10,Newman-old,Marceau}, as it simultaneously allows analytical derivations of the epidemic thresholds and {\color{black}an approximate} description of the temporal evolution of the system, besides providing a way to isolate the effects on spreading dynamics of each possible interaction mechanism, such as variations of infectivity, susceptibility or infectious periods. In addition, it enables us to solve the two paradigmatic modeling scenarios (SIS and SIR), identifying relevant differences between the two cases that arise as a consequence of disease interactions.

The model here presented and analyzed, even if it is based on an HMF and so, constitutes a first approximation to the problem that neglects dynamical correlations among nodes, is able to identify novel phenomena qualitatively different to what is found on HMF descriptions of non interacting systems. First, our model foresees different thresholds for SIR and SIS models, which arises from the asymmetry between the interaction mechanisms that take place under both models. Additionally, in what regards size scaling of epidemic thresholds, at least in the SIS case, we have identified some relevant situations that yield threshold dependences on network's size which are different from what is found in HMF models for single diseases. At this particular, it is worth mentioning that asymmetries between SIS and SIR critical properties \cite{eames,caste}, or a different behavior with respect to that predicted by the HMF for epidemic thresholds \cite{Guerra} have also been identified in the context of single diseases spreading over one network, if the HMF approach is abandoned. 

The situation here is qualitatively different, as all these divergent results with respect to classical HMF models of single diseases do not arise as a consequence of considering any dynamical correlation but as a consequence of disease-disease interactions. This has two relevant implications. On the one hand, deviations from HMF results on epidemic thresholds on single diseases identified as a consequence of considering dynamical correlations are of quantitatively residual relevance, precisely because, at the epidemic threshold, those correlations tend to vanish and, there, HMF models perform consistently well \cite{Guerra}. Instead, in our case, differences between SIS and SIR thresholds, for example, are totally different, as they even may depend on conceptually different interaction mechanisms. On the other hand, HMF is exact when dealing with annealed networks \cite{caste}, and so, on these type of networks which lack, by definition, dynamical correlations, both SIS-SIR asymmetries and eventually anomalous threshold dependences with size that we have identified in this work constitute phenomena not found before.

In addition, for the first time, the modeling framework proposed here allows to isolate the independent effects of the different mechanisms of interaction that can determine the critical properties of the model. On the one hand, this allows us to foresee that, in a SIS model, certain interaction schemes may yield to effects of one disease on another of different sign as a function of the prevalence levels of the former. On the other hand, for the SIR model, we have also discussed how the interaction between the two diseases and the different dynamical classes give rise to a richer phenomenology. In particular, we have shown that due to the transitory nature of SIR spreading processes, the moment at which the interaction of the two diseases is made effective might greatly determine the values of the epidemic threshold of the disease whose course is modified by the other one. 

Further advances in the field should address the influences of dynamical correlations between nodes on the spreading of interacting epidemics. This constitutes a truly conceptual challenge as, for example, pairwise descriptions of the disease dynamics \cite{eames}, should transform into quadruplet-based descriptions, as, in this case, dynamical correlations go beyond the dynamical state of neighbors in a network, but comprises both dynamical states on the two networks involved, which multiplies the complexity of the description.}

\appendix

\section{Epidemic thresholds on regular networks}

In the particular case in which both networks are regular graphs, an independent derivation of the epidemic threshold can be obtained by a linear stability analysis.  In that case, for which $P(k,l)=\delta(k-k_{o})\delta(l-l_{o})$, the dynamics can be described by the following system of four equations:

\begin{eqnarray*}
\dot{SS}=-k_{o}\lambda_{1}SS\cdot IS-l_{o}\lambda_{2}SS\cdot SI-k_{o}\beta_{1}^{b}\lambda_{1}SS·II-l_{o}\beta_{2}^{b}\lambda_{2}SS\cdot II+\mu_{1}IS+\mu_{2}SI,\\
\dot{IS}=k_{o}\lambda_{1}SS\cdot IS+k_{o}\beta_{1}^{b}\lambda_{1}SS\cdot II-l_{o}\beta_{2}^{a}\lambda_{2}SI\cdot IS-l_{o}\beta_{2}^{a}\beta_{2}^{b}\lambda_{2}IS\cdot II-\mu_{1}IS+\eta_{2}\mu_{2}II,\\
\dot{SI}=l_{o}\lambda_{2}SS\cdot SI+l_{o}\beta_{2}^{b}\lambda_{2}SS\cdot II-k_{o}\beta_{1}^{a}\lambda_{1}IS\cdot SI-k_{o}\beta_{1}^{a}\beta_{1}^{b}\lambda_{1}SI\cdot II-\mu_{2}SI+\eta_{1}\mu_{1}II,\\
\dot{II}=k_{o}\beta_{1}^{a}\lambda_{1}IS\cdot SI+l_{o}\beta_{2}^{a}\lambda_{2}IS\cdot SI+k_{o}\beta_{1}^{a}\beta_{1}^{b}\lambda_{1}SI\cdot II+l_{o}\beta_{2}^{a}\beta_{2}^{b}\lambda_{2}IS\cdot II-(\eta_{1}\mu_{1}+\eta_{2}\mu_{2})II,\\
\end{eqnarray*}
one of which is linearly dependent of the rest. Thus, we will only analyze the system constituted by the three last equations and use $s=1-IS-SI-II$. In order to perform a linear stability analysis we first linearize the system around the equilibrium point and we calculate the Jacobian to get:

\begin{equation}
J=\left| \begin{array}{*{3}{c}}
\lambda_{1}k_{o}-\mu_{1} & 0 & k_{o}\beta_{1}^{b}\lambda_{1}+\eta_{2}\mu_{2} \\
0 & \lambda_{2}l_{o}-\mu_{2} & l_{o}\beta_{2}^{b}\lambda_{2}+\eta_{1}\mu_{1} \\
0 & 0 & -(\eta_{1}\mu_{1}+\eta_{2}\mu_{2})
\end{array}\right|
\end{equation}
which, taking advantage of the fact that the Jacobian itself is just the product of the elements in the diagonal, yields the stability conditions detailed in table \ref{table3}.

\begin{table*}
\caption{Linear stability analysis of the fixed point $(IS,SI,II)=(0,0,0)$}
\begin{center}
\begin{tabular}{c|c|p{60mm}l}
\hline\hline
Eigen-value & Eigen-vector $(IS,SI,II)$ & Stability condition (epidemic threshold) \\\hline
$\xi_{1}=\lambda_{1}k_{o}-\mu_{1}$ & $\vec{\psi_{1}}=(1,0,0)$ & $\lambda_{1}<{\lambda_{1}^{c}}_{i}=\mu_{1}/k_{o}$ \\\hline
$\xi_{2}=\lambda_{2}l_{o}-\mu_{2}$ & $\vec{\psi_{2}}=(0,1,0)$ & $\lambda_{2}<{\lambda_{2}^{c}}_{i}=\mu_{2}/l_{o}$ \\\hline
$\xi_{3}=-(\eta_{1}\mu_{1}+\eta_{2}\mu_{2})$ & $\vec{\psi_{3}}=\left(-\frac{\eta_{2}\mu_{2}+\beta_{1}^{b}\lambda_{1}k_{o}}{(\lambda_{1}k_{o}-\mu_{1})+(\eta_{1}\mu_{1}+\eta_{2}\mu_{2})},-\frac{\eta_{1}\mu_{1}+\beta_{2}^{b}\lambda_{2}l_{o}}{(\lambda_{2}l_{o}-\mu_{2})+(\eta_{1}\mu_{1}+\eta_{2}\mu_{2})},1\right)$ & always stable \\\hline
\end{tabular}
\end{center}
\label{table3}
\end{table*}
 
If we look carefully at the results sketched in table \ref{table3},  we firstly recognize the appearance of a couple of critical values for the infectiousnesses ${\lambda_{1}^{c}}_{i}$ and ${\lambda_{2}^{c}}_{i}$ which will be referred to in what follows as the primary thresholds of their respective diseases, just like in the general case described in the main text. These threshold values stand for the \emph{minimum values of the infectiousnesses that yield epidemic outbreaks after the introduction of infinitesimal seeds of infected individuals of each disease in an initially healthy population}. Therefore, the condition $\lambda_{1}>{\lambda_{1}^{c}}_{i}$ must be verified in order to have an epidemic outbreak for the first disease once an infinitesimal seed $(IS,SI,II)=(\epsilon,0,0)$ has been introduced on a system being at the disease-free fixed point $(IS,SI,II)=(0,0,0)$. On the other hand, the condition ${\lambda_{2}^{c}}_{i}$ plays an equivalent role for the second disease with respect to a seed $(IS,SI,II)=(0,\epsilon,0)$. 

The fixed point $(IS,SI,II)=(0,0,0)$ is, strictly speaking, not the only possible disease-free fixed point in our model. Other two partially disease-free fixed points can exist: a first fixed point in which disease 1 is installed in the system at a certain prevalence $\pi_{1}$ whilst disease 2 is absent $(IS,SI,II)=(\pi_{1},0,0)$ and its cognate $(IS,SI,II)=(0,\pi_{2},0)$, for which there is no individual infected with disease 1. As we are going to show, the stability of these fixed points depends on the prevalence fractions $\pi_{1}$ and $\pi_{2}$, which are the stationary proportions of sick individuals of each disease:

\begin{equation}
\pi_{1}=\frac{\lambda_{1}k_{o}-\mu_{1}}{\lambda_{1}k_{o}}
\label{pi1}
\end{equation}

\begin{equation}
\pi_{2}=\frac{\lambda_{2}l_{o}-\mu_{2}}{\lambda_{2}l_{o}}
\label{pi2}
\end{equation}

Considering that, let us study the stability of the first fixed point $(IS,SI,II)=(\pi_{1}=\frac{\lambda_{1}k_{o}-\mu_{1}}{\lambda_{1}k_{o}},0,0)$ as a function of $\lambda_{1}$: the Jacobian, around this fixed point takes the following form:

\begin{equation}
J=\left| \begin{array}{*{3}{c}}
(\mu_{1}-\lambda_{1}k_{o}) & (\mu_{1}-\lambda_{1}k_{o})(1+\beta_{2}^{a}\frac{\lambda_{2}l_{o}}{\lambda_{1}k_{o}}) & \beta_{1}^{b}\lambda_{1}k_{o}+\eta_{2}\mu_{2}+(\mu_{1}-\lambda_{1}k_{o})(\beta_{1}^{b}+1+\beta_{2}^{a}\beta_{2}^{b}\frac{\lambda_{2}l_{o}}{\lambda_{1}k_{o}})\\
0 & (\lambda_{2}l_{o}-\mu_{2}) +(\mu_{1}-\lambda_{1}k_{o})(\frac{\lambda_{2}l_{o}}{\lambda_{1}k_{o}}+\beta_{1}^{a})& \beta_{2}^{b}\mu_{1}\frac{\lambda_{2}l_{o}}{\lambda_{1}k_{o}}+\eta_{1}\mu_{1} \\
0 & (\lambda_{1}k_{o}-\mu_{1})(\beta_{2}^{a}\frac{\lambda_{2}l_{o}}{\lambda_{1}k_{o}}+\beta_{1}^{a}) & (\lambda_{1}k_{o}-\mu_{1})\beta_{2}^{a}\beta_{2}^{b}\frac{\lambda_{2}l_{o}}{\lambda_{1}k_{o}}-(\eta_{1}\mu_{1}+\eta_{2}\mu_{2})
\end{array}\right|
\label{Jacobian}
\end{equation}

By solving the equation derived from imposing that $J=0$, we obtain the values of the parameters leading to stability inversion. As it can be shown at naked eye,  $(\mu_{1}-\lambda_{1}k_{o})$ is the eigenvalue $\xi_{1}$ associated to the eigenvector $\vec{\psi_{1}}=(1,0,0)$, and the condition $\xi_{1}=0$ yields again the same condition $\lambda_{1}=\mu_{1}/k_{o}$, thus defining the threshold of the classical SIS model. Regarding the other two eigenvalues $\xi_{2}$ and $\xi_{3}$, the result is more cumbersome. Despite of that, the vanishing of the $2$x$2$ determinant of the right-inferior corner of the Jacobian matrix \ref{Jacobian} yields:  

\begin{equation}
\lambda_{2}={\lambda_{2}^{c}}_{ii}=\frac{\lambda_{1}k_{o}}{l_{o}}\frac{\eta_{2}\mu_{2}((\beta_{1}^{a}(\lambda_{1}k_{o}-\mu_{1})+\mu_{2})+\eta_{1}\mu_{1}\mu_{2}}{\mu_{1}(\eta_{1}\mu_{1}+\eta_{2}\mu_{2})+(\lambda_{1}k_{o}-\mu_{1})(\beta_{1}^{a}\beta_{2}^{a}\beta_{2}^{b}(\lambda_{1}k_{o}-\mu_{1})+\beta_{2}^{a}\beta_{2}^{b}\mu_{2}+\eta_{1}\mu_{1}\beta_{2}^{a}+\beta_{1}^{a}\beta_{2}^{b}\mu_{1})},
\label{SecondaryThreshold2}
\end{equation}
to which the general expression presented in the main text for the epidemic threshold reduces when $P(k,l)=\delta(k-k_{o})\delta(l-l_{o})$. The agreement between numerical simulations and the analytic expression of the threshold presented here is only accurate when $\lambda_{1}k_{o}\simeq\mu_{1}$, and the reason is easily understandable. Let us compare the reaction of the system to the introduction of a small seed of  $SI$ individuals when both diseases are absent ($(IS,SI,II)=(0,0,0)$, case 1) or when the first disease was already installed in the system ($(IS,SI,II)=(\pi_{1},0,0)$, case 2). As we have argued in the precedent sections, the epidemic threshold is different in each case, and the reason is simply the presence, in the second case, of a fraction $\pi_{1}$ of $IS$ of individuals for which the infectiousness and recovery rates for disease 2 are different with respect to the rest of individuals. As a consequence, the mean values of the dynamical parameters averaged over the whole population $\langle\lambda_{2}\rangle$ and $\langle\mu_{2}\rangle$, are different in both cases and will yield different values for the epidemic thresholds. Thus, in order to accurately evaluate the secondary threshold for disease 2 it is essential to know with enough precision the prevalence $\pi_{1}$ corresponding to a single SIS model, as a function of $\lambda_{1}$. The problem arises from the fact that, precisely, the derivation of \ref{SecondaryThreshold2} explicitly assumes that the bijection between $\pi_{1}$ and $\lambda_{1}$ is governed by the mean field stationary expression \ref{pi1} which is only precise when $\lambda_{1}k_{o}\simeq\mu_{1}$ \cite{EPLnostre}. In fact, a way to rebuild a more accurate secondary threshold curve can be achieved if, from equation \ref{pi1} we substitute $\lambda_{1}$ as a function of $\pi_{1}$, and then introduce the so obtained expression into \ref{SecondaryThreshold2} to get:

\begin{equation}
{\lambda_{2}^{c}}_{ii}(\pi_{1})=\frac{1}{l_{o}}\frac{\eta_{2}\mu_{2}(\beta_{1}^{a}\frac{\mu_{1}\pi_{1}}{(1-\pi_{1})}+\mu_{2})+\eta_{1}\mu_{1}\mu_{2}}{(1-\pi_{1})(\eta_{1}\mu_{1}+\eta_{2}\mu_{2})+\pi_{1}(\beta_{1}^{a}\beta_{2}^{a}\beta_{2}^{b}\frac{\mu_{1}\pi_{1}}{(1-\pi_{1})}+\beta_{2}^{a}\beta_{2}^{b}\mu_{2}+\eta_{1}\mu_{1}\beta_{2}^{a}+\beta_{1}^{a}\beta_{2}^{b}\mu_{1})}
\label{SecondaryThreshold2*}
\end{equation}
which allows to evaluate directly the threshold as a function of $\pi_{1}$ rather than of $\lambda_{1}$. Thus, by introducing in equation \ref{SecondaryThreshold2*} the $\pi_{1}$ values obtained from the simulations instead of the theoretical prediction of the mean field (equation \ref{pi1}), we recover the curves for the secondary threshold represented with red lines in Figures \ref{fig2} and \ref{fig3} of the main text, in quantitative agreement with results from simulations. 

Obviously, the same arguments stand for the secondary threshold of the first disease, which obeys the following expression:

\begin{equation}
{\lambda_{1}^{c}}_{ii}(\pi_{2})=\frac{1}{k_{o}}\frac{\eta_{1}\mu_{1}(\beta_{2}^{a}\frac{\mu_{2}\pi_{2}}{(1-\pi_{2})}+\mu_{1})+\eta_{2}\mu_{2}\mu_{1}}{(1-\pi_{2})(\eta_{2}\mu_{2}+\eta_{1}\mu_{1})+\pi_{2}(\beta_{2}^{a}\beta_{1}^{a}\beta_{1}^{b}\frac{\mu_{2}\pi_{2}}{(1-\pi_{2})}+\beta_{1}^{a}\beta_{1}^{b}\mu_{1}+\eta_{2}\mu_{2}\beta_{1}^{a}+\beta_{2}^{a}\beta_{1}^{b}\mu_{2})}
\label{SecondaryThreshold1*}
\end{equation}
When the networks are heterogeneous, this reformulation of the threshold curves as a function of $\pi_{1}$ or $\pi_{2}$ can not be done straightforwardly as no analytical bijection $\lambda_{1}(\pi_{1})$ or $\lambda_{2}(\pi_{2})$ can be reached. The best we can do is to use a numerically built-up relationship $\lambda_{1}(\theta_{1})$ (or $\lambda_{2}(\theta_{2})$), and introduce it into the general expression for the threshold. Although the accuracy of the curves for the secondary threshold is very satisfactory in Fig. \ref{figure4}, we identify this effect to be the source of the slight divergence between the analytical and numerical secondary thresholds shown in Fig. \ref{figure6}.

\section{Vanishing conditions for epidemic thresholds}

Here we provide a systematic analysis of the behavior of the secondary threshold in our model for large scale-free networks. In particular, we inspect whether the coupling between the spreading of the two diseases in the terms described in our model can modify the classical, single-disease scheme, and, if so, under what conditions. The epidemic threshold for the first disease in our model reads as:

\begin{equation}
\lambda_{1}^{c}(\sigma_{2})=\mu_{1}\frac{\langle k \rangle}
{
	\sum_{k,l}
		{
			P(k,l)\frac
						{
							k^{2}l^{2}\sigma_{2}^{2}\beta_{2}^{a}\beta_{1}^{a}\beta_{1}^{b}
							+
							lk^{2}\sigma_{2}
												(
												\eta_{2}\mu_{2}\beta_{1}^{a}
												+
												\beta_{1}^{b}
												                     (
												                       \beta_{1}^{a}\mu_{1}
												                       +
												                       \beta_{2}^{a}\mu_{2}
												                      )
												 )
							+
							k^{2}\mu_{2}
							                     (
							                       \eta_{1}\mu_{1}
							                       +
							                       \eta_{2}\mu_{2}
							                     )
						}
						{
							l^{2}\sigma_{2}^{2}\beta_{2}^{a}\eta_{1}
							+
							l\sigma_{2}
												(
												\eta_{1}\mu_{1}+\eta_{2}\mu_{2}+ \beta_{2}^{a}\eta_{1}\mu_{2}
											                                        )
							+
							\mu_{2}
							                     (
							                       \eta_{1}\mu_{1}	
							                       +
							                       \eta_{2}\mu_{2}
							                     )
				}
					}
		}
\label{umbral}
\end{equation} 
and we have to address its behavior in the limit $N\rightarrow\infty$. Substituting sums by integrals in Eq. (\ref{umbral}), one gets:

\begin{equation}
\lim_{N\rightarrow\infty}{\lambda_{1}^{c}(\sigma_{2})}=
\lim_{(k_{max},l_{max})\rightarrow\infty}
{
\mu_{1}\frac{\int_{k_{min}}^{k_{max}}{\int_{l_{min}}^{l_{max}}{P(k,l)k\ dk\ dl}}}
{
	\int_{k_{min}}^{k_{max}}
		{\int_{l_{min}}^{l_{max}}
		{
			P(k,l)\frac
						{
							k^{2}l^{2}\sigma_{2}^{2}\beta_{2}^{a}\beta_{1}^{a}\beta_{1}^{b}
							+
							lk^{2}\sigma_{2}
												(
												\eta_{2}\mu_{2}\beta_{1}^{a}
												+
												\beta_{1}^{b}
												                     (
												                       \beta_{1}^{a}\mu_{1}
												                       +
												                       \beta_{2}^{a}\mu_{2}
												                      )
												 )
							+
							k^{2}\mu_{2}
							                     (
							                       \eta_{1}\mu_{1}
							                       +
							                       \eta_{2}\mu_{2}
							                     )
						}
						{
							l^{2}\sigma_{2}^{2}\beta_{2}^{a}\eta_{1}
							+
							l\sigma_{2}
												(
												\eta_{1}\mu_{1}+\eta_{2}\mu_{2}+ \beta_{2}^{a}\eta_{1}\mu_{2}
											                                        )
							+
							\mu_{2}
							                     (
							                       \eta_{1}\mu_{1}	
							                       +
							                       \eta_{2}\mu_{2}
							                     )
				}
					}
		}dk\ dl
		}
		}.
\label{umbralints}
\end{equation} 
To study the behavior of the threshold in the thermodynamic limit, we present here an analysis which is essentially based on the following result, whose proof is an exercise of elementary algebra that we present for the sake of completeness in the last appendix. Given two polynomials $P(k)$, $Q(k)$, we have that:
\begin{equation}
{\left|\lim_{k_{max}\rightarrow \infty}{\int_{k_{min}}^{k_{max}}{  \frac{P(k)}{Q(k)}dk  }  }\right|=\infty \leftrightarrow deg(Q)-deg(P)\leq1
}
\label{theorem}
\end{equation}
Focusing on scale-free connectivity distributions, we will distinguish two different scenarios in this section: uncorrelated and totally correlated layers. In the first case, both networks present a scale free distribution in which the connectivity of a node in a layer is essentially independent of its degree on the other layer, in such a way that the composed connectivity distribution verifies $P(k,l)=C_{o}k^{-\gamma}l^{-\Gamma}$. Instead, in the second case, although both layers are also scale free networks, the degree of any given node in both layers is forced to be the same. Therefore, nodes which are hubs in a layer are so in the other, and the composed degree distribution is $P(k,l)=C_{o}\delta(k-l)k^{-\gamma}$. In addition, we assume that both $\gamma$ and $\Gamma$ exponents are rational, hence, we can write $\gamma=w/x$ and $\Gamma=y/z$ with $(w,x,y,z)\in \mathbb{N}$. By addressing these two opposite scenarios, our aim is to characterize the difference, in terms of the spreading dynamics, between the coupling of two diseases that spread by independent means --which will give place to different networks of contacts-- and the coupling of two related diseases --or variations of the same disease-- that spread following the very same mechanisms and, thus, they do so over highly correlated networks of contacts.

\subsection{Uncorrelated scale free layers}

If we substitute $P(k,l)=C_{o}k^{-\gamma}l^{-\Gamma}$ into eq. \ref{umbralints}, we can factorize the double integrals into independent terms:

\begin{equation}
\lim_{N\rightarrow\infty}{\lambda_{1}^{c}(\sigma_{2})}=
\lim_{(k_{max},l_{max})\rightarrow\infty}
{
\mu_{1}\frac{\int_{k_{min}}^{k_{max}}{k^{1-\gamma}dk} \int_{l_{min}}^{l_{max}}{l^{-\Gamma}dl}}
{
	\int_{k_{min}}^{k_{max}}
		{k^{2-\gamma}dk}
			\int_{l_{min}}^{l_{max}}
		{
			\frac
						{
							l^{2}\sigma_{2}^{2}\beta_{2}^{a}\beta_{1}^{a}\beta_{1}^{b}
							+
							l\sigma_{2}
												(
												\eta_{2}\mu_{2}\beta_{1}^{a}
												+
												\beta_{1}^{b}
												                     (
												                       \beta_{1}^{a}\mu_{1}
												                       +
												                       \beta_{2}^{a}\mu_{2}
												                      )
												 )
							+
							\mu_{2}
							                     (
							                       \eta_{1}\mu_{1}
							                       +
							                       \eta_{2}\mu_{2}
							                     )
						}
						{
							l^{\Gamma}\left(l^{2}\sigma_{2}^{2}\beta_{2}^{a}\eta_{1}
							+
							l\sigma_{2}
												(
												\eta_{1}\mu_{1}+\eta_{2}\mu_{2}+ \beta_{2}^{a}\eta_{1}\mu_{2}
											                                        )
							+
							\mu_{2}
							                     (
							                       \eta_{1}\mu_{1}	
							                       +
							                       \eta_{2}\mu_{2}
							                     )\right)
				}
		}dl
		}
		}
\label{uncorr}
\end{equation} 
The condition for the numerator to diverge is $\gamma\leq2\vee\Gamma\leq1$. These conditions are verified by networks whose mean connectivity diverges in the thermodynamic limit. For this reason, networks with those small exponents neither are reasonable systems to be used to describe any information diffusion process over them nor are found in real systems. Taking this into account, we will restrict our analysis to the more epidemiologically relevant scenario in which $\gamma>2$ and $\Gamma>2$, although our reasoning is generalizable to any value of the exponents. 

Thus, in the scenario $\gamma>2$ and $\Gamma>2$, the numerator remains always finite and the only phenomenon of interest that could be found is an eventual vanishing of the threshold due to a divergence in the denominator --$D(k,l)$ in what follows--. $D(k,l)$ can be factorized to give $D(k,l)=D_{1}(k)D_{2}(l)$, as it can be seen from Eq. \ref{uncorr}. The integral in $k$, $D_{1}(k)$ will diverge if and only if $\gamma\leq3$, but the integral in $l$, $D_{2}(l)$, might independently diverge under some conditions. That situation would suppose the vanishing of the threshold of the first disease as a consequence of its coupling to the second rather than to internal, dynamical or topological features. To find out the conditions for $D_{2}(l)$, lets make the following change of variable:
\begin{equation}
m=l^{1/z} 
\end{equation}
so as to change the dependence of the denominator $D(k,l)=D_{1}(k)D_{2}(l)$ in Eq. \ref{uncorr} to $D(k,m)=D_{1}(k)D_{2}(m)$: 
\begin{equation}
\lim_{N\rightarrow\infty}{\lambda_{1}^{c}(\sigma_{2})}=
\lim_{(k_{max},l_{max})\rightarrow\infty}
{
\mu_{1}\frac{\int_{k_{min}}^{k_{max}}{k^{1-\gamma}dk}}
{
	\int_{k_{min}}^{k_{max}}
		{k^{2-\gamma}dk}
			\int_{l_{min}^{1/z}}^{l_{max}^{1/z}}
		{
			\frac
						{
							\left(m^{2z}\sigma_{2}^{2}\beta_{2}^{a}\beta_{1}^{a}\beta_{1}^{b}
							+
							m^{z}\sigma_{2}
												(
												\eta_{2}\mu_{2}\beta_{1}^{a}
												+
												\beta_{1}^{b}
												                     (
												                       \beta_{1}^{a}\mu_{1}
												                       +
												                       \beta_{2}^{a}\mu_{2}
												                      )
												 )
							+
							\mu_{2}
							                     (
							                       \eta_{1}\mu_{1}
							                       +
							                       \eta_{2}\mu_{2}
							                     )\right)zm^{z-1}
						}
						{
							m^{y}\left(m^{2z}\sigma_{2}^{2}\beta_{2}^{a}\eta_{1}
							+
							m^{z}\sigma_{2}
												(
												\eta_{1}\mu_{1}+\eta_{2}\mu_{2}+ \beta_{2}^{a}\eta_{1}\mu_{2}
											                                        )
							+
							\mu_{2}
							                     (
							                       \eta_{1}\mu_{1}	
							                       +
							                       \eta_{2}\mu_{2}
							                     )\right)
				}
		}dm
		}
		}
\label{uncorr2}
\end{equation} 
The last expression has the advantage that the argument of the integral in $m$, $D_{2}(m)$ is just the quotient of two polynomials, and thus, its behavior in the limit $\l_{max}\rightarrow\infty$ is governed by Eq. (\ref{theorem}) and depends only on the difference of degrees of the denominator and the numerator. If 
\begin{equation}
D_{2}(m)=\int_{l_{min}^{1/z}}^{l_{max}^{1/z}}
		{
			\frac
						{
							\left(m^{2z}\sigma_{2}^{2}\beta_{2}^{a}\beta_{1}^{a}\beta_{1}^{b}
							+
							m^{z}\sigma_{2}
												(
												\eta_{2}\mu_{2}\beta_{1}^{a}
												+
												\beta_{1}^{b}
												                     (
												                       \beta_{1}^{a}\mu_{1}
												                       +
												                       \beta_{2}^{a}\mu_{2}
												                      )
												 )
							+
							\mu_{2}
							                     (
							                       \eta_{1}\mu_{1}
							                       +
							                       \eta_{2}\mu_{2}
							                     )\right)zm^{z-1}
						}
						{
							m^{y}\left(m^{2z}\sigma_{2}^{2}\beta_{2}^{a}\eta_{1}
							+
							m^{z}\sigma_{2}
												(
												\eta_{1}\mu_{1}+\eta_{2}\mu_{2}+ \beta_{2}^{a}\eta_{1}\mu_{2}
											                                        )
							+
							\mu_{2}
							                     (
							                       \eta_{1}\mu_{1}	
							                       +
							                       \eta_{2}\mu_{2}
							                     )\right)
				}
		}dm=\int_{l_{min}^{1/z}}^{l_{max}^{1/z}}
		{
			\frac
						{P(m)}{Q(m)}
						}dm
\label{D2}
\end{equation} 
we have that --in the general case in which none of the $\beta$ and $\eta$ parameters vanishes-- $deg(Q)-deg(P)=1+y-z$, and hence, the conditions for $D_{2}(m)$ to diverge are
\begin{equation}
\lim_{(k_{max},l_{max})\rightarrow\infty}
{
\int_{l_{min}^{1/z}}^{l_{max}^{1/z}}
		{
			\frac
						{P(m)}{Q(m)}
						}dm=\infty \leftrightarrow y-z+1\leq 1 \leftrightarrow \Gamma=y/z\leq 1
						}
						\label{nonulos}
\end{equation}
Therefore, in the region of interest $\gamma>2\wedge\Gamma>2$, the factor $D_{2}(m)$ can not diverge and make the threshold vanishes. It is worth noticing that the condition $\Gamma<1$ does not guarantee the vanishing of the threshold, as it also makes the numerator to diverge. Expression \ref{nonulos} is valid for the case in which none of the  $\beta$ and $\eta$ parameters of the model vanishes. If this is not the case, i.e., if some of the infectiousness variations $\beta$ do vanish $-$the rest of the parameters can not do that within a realistic epidemiological framework$-$, the degrees of the numerator and the denominator in Eq. \ref{D2} could vary. In table \ref{tablenocorr} we systematically address  all possible combinations of null parameters, and the composed conditions yielding a vanishing threshold for each case.

\begin{table*}
\caption{Divergence conditions for $D_{2}(m)$ for disease 1. Case 1: uncorrelated scale free layers $P(k,l)=C_{o}k^{-\gamma}l^{-\Gamma}$, $\gamma=(w/x)$, $\Gamma=(y/z)$.}
\begin{center}
\begin{tabular}{c|c|cl}
\hline\hline
\begin{tabular}{c} $(\beta_{1}^{a},\beta_{1}^{b},\beta_{2}^{a})$\\ parameters\\  \end{tabular} & \begin{tabular}{c} $deg(Q)-deg(P)$\\ in $D_{2}(m)=P(m)/Q(m)$\\  \end{tabular}  & \begin{tabular}{c} Divergence condition\\  \end{tabular}\\\hline
\begin{tabular}{c} $(\beta_{1}^{a},\beta_{1}^{b},\beta_{2}^{a})$\\ $(\beta_{1}^{a},\beta_{1}^{b},0)$\\ $(\beta_{1}^{a},0,0)$\\  \end{tabular}
 & $y-z+1$ & $\gamma\leq3 \vee \Gamma\leq1$ \\\hline
\begin{tabular}{c} $(0,\beta_{1}^{b},\beta_{2}^{a})$\\ $(\beta_{1}^{a},0,\beta_{2}^{a})$\\ $(0,\beta_{1}^{b},0)$\\ $(0,0,0)$\\ \end{tabular}
 & $y+1$ & $\gamma\leq3 \vee \Gamma\leq0$ \\\hline
$(0,0,\beta_{2}^{a})$ & $y+z+1$ & $\gamma\leq3 \vee \Gamma\leq-1$ \\\hline
\end{tabular}
\end{center}
\label{tablenocorr}
\end{table*}

Once again, we see that, whatever the interacting scheme between both diseases is, the denominator remains always finite provided that $\Gamma>2$. In conclusion, if no degree correlation is introduced between layers, and for realistic systems characterized by double power laws verifying $\gamma>2\wedge\Gamma>2$, the behavior of the thresholds in the thermodynamic limit is essentially the same of the uncoupled systems: the threshold associated to the first disease vanishes if and only if the exponent of its own network verifies $\gamma\leq3$, whatever the exponent of the second network. The coupling will introduce, in general, only a finite pre-factor. The symmetric situation obviously stands for the threshold of the second disease. 

\subsection{Totally correlated scale free layers}

If we consider the case in which $P(k,l)=C_{o}\delta(k-l)k^{-\gamma}$, where $\gamma=w/x$ with $(w,x) \in \mathbb{N}$, the epidemic threshold reads as:
\begin{equation}
\lim_{N\rightarrow\infty}{\lambda_{1}^{c}(\sigma_{2})}=
\lim_{(k_{max},l_{max})\rightarrow\infty}
{
\mu_{1}\frac{\int_{k_{min}}^{k_{max}}{k^{1-\gamma}dk}}
{
			\int_{k_{min}}^{k_{max}}
		{
			\frac
						{
							k^{2}\sigma_{2}^{2}\beta_{2}^{a}\beta_{1}^{a}\beta_{1}^{b}
							+
							k\sigma_{2}
												(
												\eta_{2}\mu_{2}\beta_{1}^{a}
												+
												\beta_{1}^{b}
												                     (
												                       \beta_{1}^{a}\mu_{1}
												                       +
												                       \beta_{2}^{a}\mu_{2}
												                      )
												 )
							+
							\mu_{2}
							                     (
							                       \eta_{1}\mu_{1}
							                       +
							                       \eta_{2}\mu_{2}
							                     )
						}
						{
							k^{\gamma-2}\left(k^{2}\sigma_{2}^{2}\beta_{2}^{a}\eta_{1}
							+
							k\sigma_{2}
												(
												\eta_{1}\mu_{1}+\eta_{2}\mu_{2}+ \beta_{2}^{a}\eta_{1}\mu_{2}
											                                        )
							+
							\mu_{2}
							                     (
							                       \eta_{1}\mu_{1}	
							                       +
							                       \eta_{2}\mu_{2}
							                     )\right)
				}
		}dk
		}
		}
\label{corr}
\end{equation} 
hence, we do not recover the factorization of the denominator previously observed. Instead, after changing the variable to $m=k^{1/x}$, we get the following expression:
\begin{equation}
\lim_{N\rightarrow\infty}{\lambda_{1}^{c}(\sigma_{2})}=
\lim_{(k_{max})\rightarrow\infty}
{
\mu_{1}\frac{\int_{k_{min}}^{k_{max}}{k^{1-\gamma}dk}}
{
			\int_{k_{min}^{1/x}}^{k_{max}^{1/x}}
		{
			\frac
						{
							\left(m^{2x}\sigma_{2}^{2}\beta_{2}^{a}\beta_{1}^{a}\beta_{1}^{b}
							+
							m^{x}\sigma_{2}
												(
												\eta_{2}\mu_{2}\beta_{1}^{a}
												+
												\beta_{1}^{b}
												                     (
												                       \beta_{1}^{a}\mu_{1}
												                       +
												                       \beta_{2}^{a}\mu_{2}
												                      )
												 )
							+
							\mu_{2}
							                     (
							                       \eta_{1}\mu_{1}
							                       +
							                       \eta_{2}\mu_{2}
							                     )\right)xm^{x-1}
						}
						{
							m^{w-2x}\left(m^{2x}\sigma_{2}^{2}\beta_{2}^{a}\eta_{1}
							+
							m^{x}\sigma_{2}
												(
												\eta_{1}\mu_{1}+\eta_{2}\mu_{2}+ \beta_{2}^{a}\eta_{1}\mu_{2}
											                                        )
							+
							\mu_{2}
							                     (
							                       \eta_{1}\mu_{1}	
							                       +
							                       \eta_{2}\mu_{2}
							                     )\right)
				}
		}dm
		}
		}
\label{corr2}
\end{equation} 
In this expression, the numerator diverges for $\gamma\leq 2$. In turn, the denominator:
\begin{equation}
D(m)=
			\int_{k_{min}^{1/x}}^{k_{max}^{1/x}}
		{
			\frac
						{
							\left(m^{2x}\sigma_{2}^{2}\beta_{2}^{a}\beta_{1}^{a}\beta_{1}^{b}
							+
							m^{x}\sigma_{2}
												(
												\eta_{2}\mu_{2}\beta_{1}^{a}
												+
												\beta_{1}^{b}
												                     (
												                       \beta_{1}^{a}\mu_{1}
												                       +
												                       \beta_{2}^{a}\mu_{2}
												                      )
												 )
							+
							\mu_{2}
							                     (
							                       \eta_{1}\mu_{1}
							                       +
							                       \eta_{2}\mu_{2}
							                     )\right)xm^{x-1}
						}
						{
							m^{w-2x}\left(m^{2x}\sigma_{2}^{2}\beta_{2}^{a}\eta_{1}
							+
							m^{x}\sigma_{2}
												(
												\eta_{1}\mu_{1}+\eta_{2}\mu_{2}+ \beta_{2}^{a}\eta_{1}\mu_{2}
											                                        )
							+
							\mu_{2}
							                     (
							                       \eta_{1}\mu_{1}	
							                       +
							                       \eta_{2}\mu_{2}
							                     )\right)
				}
		}dm
		=\int_{k_{min}^{1/x}}^{k_{max}^{1/x}}
		{
			\frac
						{P(m)}{Q(m)}
						}dm
\label{D2corr}
\end{equation} 
diverges, according to equation \ref{theorem}, if and only if $deg(Q)-deg(P)\leq1$. In the general case in which none of the $\beta$ parameters vanishes, we have that $deg(Q)-deg(P)=w-3x+1$, and thus:
\begin{equation}
\lim_{N\rightarrow\infty}{\lambda_{1}^{c}(\sigma_{2})}=0 \leftrightarrow \gamma\leq3
\label{nonull2}
\end{equation}
As in the previous case, different combinations of null $\beta$ parameters can change this result, as it can be seen in Table \ref{tablecorr}.
\begin{table*}
\caption{Divergence conditions for $D_{2}(m)$ for disease 1. Case 2: totally correlated scale free layers $P(k,l)=C_{o}\delta(k-l)k^{-\gamma}$, $\gamma=(w/x)$.}
\begin{center}
\begin{tabular}{c|c|cl}
\hline\hline
\begin{tabular}{c} $(\beta_{1}^{a},\beta_{1}^{b},\beta_{2}^{a})$\\ parameters\\  \end{tabular} & \begin{tabular}{c} $deg(Q)-deg(P)$\\ in $D_{2}(m)=P(m)/Q(m)$\\  \end{tabular}  & \begin{tabular}{c} Divergence condition\\  \end{tabular}\\\hline
\begin{tabular}{c} $(\beta_{1}^{a},\beta_{1}^{b},\beta_{2}^{a})$\\ $(\beta_{1}^{a},\beta_{1}^{b},0)$\\ $(\beta_{1}^{a},0,0)$\\  \end{tabular}
 & $w-3x+1$ & $\gamma\leq3$\\\hline
\begin{tabular}{c} $(0,\beta_{1}^{b},\beta_{2}^{a})$\\ $(\beta_{1}^{a},0,\beta_{2}^{a})$\\ $(0,\beta_{1}^{b},0)$\\ $(0,0,0)$\\ \end{tabular}
 & $w-2x+1$ & $\gamma\leq2$\\\hline
$(0,0,\beta_{2}^{a})$ & $w-x+1$ & $\gamma\leq1$ \\\hline
\end{tabular}
\end{center}
\label{tablecorr}
\end{table*}
Here, the phenomenology is remarkable different for all the cases in which at least one of the two variations of disease 1 infectiousness, $\beta_{1}^{a}$ or $\beta_{1}^{b}$, cancel (except for the case $\beta_{2}^{a}=\beta_{1}^{b}=0$ but $\beta_{1}^{a}\neq0$, for which eq. \ref{nonull2} also stands). In those cases, $\theta>0$ guarantees that, provided that $\gamma>2$, no threshold vanishing is observed in the thermodynamic limit, even when the exponent is within the interval $2<\gamma\leq3$. In the case in which $\beta_{1}^{a}=\beta_{1}^{b}=\beta_{2}^{a}=0$, the situation will be reciprocal, and thus, the epidemic threshold of the second disease will not vanish either for $\gamma>2$. 

\section{Proof of eq. \ref{theorem}}

Given two polynomials $P(k)$, $Q(k)$, we have to prove that
\begin{equation}
{\left|\lim_{k_{max}\rightarrow \infty}{\int_{k_{min}}^{k_{max}}{  \frac{P(k)}{Q(k)}dk  }  }\right|=\infty \leftrightarrow deg(Q)-deg(P)\leq1
}
\label{theorembis}
\end{equation}
First, we have that, if $deg(P)\geq deg(Q)$ the integral diverges, as the argument of the integral can be expressed in that case as:
\begin{equation}
{\frac{P(k)}{Q(k)}=C(k)+\frac{P'(k)}{Q'(k)}}
\label{cociente}
\end{equation}
with $deg(P')<deg(Q')$ and $deg(C)\geq0$. When substituting the last expression into eq.\ref{theorem}, the integral in $C(k)$ automatically diverges. Therefore, to prove this result for the case in which $deg(P)<deg(Q)$, let us consider the general decomposition of $Q(k)$: 
\begin{equation}
{
Q(k)=Q_{o}\prod_{i=1}^{i^{*}}{(k-a_{i})^{n_{i}}}\prod_{j=1}^{j^{*}}{(k^{2}+b_{j}k+c_{j})^{m_{j}}}
}
\label{Q}
\end{equation}
where $Q_{o}$ is a constant, $k_{min}>0$ and $b_{j}^{2}-4c_{j}<0$ $\forall j$. The values $i^{*}$ and $j^{*}$ stand for the number of different elemental factors of first and second order, respectively. So, the degree of the polynomial reads as follows:
\begin{equation}
{
deg(Q)=\sum_{i=1}^{i^{*}}{n_{i}}+2\sum_{j=1}^{j^{*}}{m_{j}},
}
\label{DegQ}
\end{equation}
where $n_{i}$ and $m_{j}$ denote the multiplicity of each of the factors of first and second order, respectively. The factorization of $Q(k)$ yields the following decomposition of the quotient $P(k)/Q(k)$ into partial fractions:
\begin{equation}
{
\int_{k_{min}}^{k_{max}}{\frac{P(k)}{Q(k)}dk}=
\sum_{i=1}^{i^{*}}{\sum_{n=1}^{n_{i}}{\int_{k_{min}}^{k_{max}}
{\frac{A_{i,n}}{(k+a_{i})^{n}}}dk}}+\sum_{j=1}^{j^{*}}{\sum_{m=1}^{m_{j}}{\int_{k_{min}}^{k_{max}}
{\frac{B_{j,m}k+C_{j,m}}{ (k^{2}+b_{j}k+c_{j})^{m}  }}dk}}
}
\label{Parfra}
\end{equation}
where the coefficients $A_{i,n},B_{j,m},C_{j,m}$ have to be calculated by expanding this expression into a single fraction, and comparing the coefficients of the numerator obtained with those of $P'(k)=P(k)/Q_{o}$. With the aim of understanding what are the conditions that make these partial integrals to diverge, let us analyze them term by term. The first sum yields logarithmic and rational functions:
\begin{equation}
{
\sum_{i=1}^{i^{*}}{\sum_{n=1}^{n_{i}}{\int_{k_{min}}^{k_{max}}
{\frac{A_{i,n}}{(k+a_{i})^{n}}}dk}}=\sum_{i=1}^{i^{*}}{A_{i,1}\ln \left(\frac{k_{max}+a_{i}}{k_{min}+a_{i}}\right)}+\sum_{i=1|n_{i}>1}^{i^{*}}{\sum_{n=2}^{n_{i}}{\frac{-A_{i,n}}{(n-1)}\left(\frac{1}{(k_{min}+a_{i})^{n-1}}-\frac{1}{(k_{max}+a_{i})^{n-1}}\right)}
}}
\label{suma}
\end{equation}
and, in the limit $k_{max}\rightarrow\infty$, only the logarithmic term diverges:
\begin{equation}
{
\lim_{k_{max}\rightarrow\infty}{\sum_{i=1}^{i^{*}}{\sum_{n=1}^{n_{i}}{\int_{k_{min}}^{k_{max}}
{\frac{A_{i,n}}{(k+a_{i})^{n}}}dk}}}=\sum_{i=1}^{i^{*}}{A_{i,1}\ln {(k_{max})}}+\vartheta}
\label{log1}
\end{equation}
where $\vartheta$ stands for a finite term, negligible when compared to $\ln(k_{max})$. On the other hand, the second sum in eq. \ref{Parfra} can be rewritten as: 
{\scriptsize 
\begin{equation}
{
\sum_{j=1}^{j^{*}}{\sum_{m=1}^{m_{j}}{\int_{k_{min}}^{k_{max}}
{\frac{B_{j,m}k+C_{j,m}}{ (k^{2}+b_{j}k+c_{j})^{m}  }dk}}}=\sum_{j=1}^{j^{*}}{\sum_{m=1}^{m_{j}}{\frac{B_{j,m}}{2}\int_{k_{min}}^{k_{max}}
{
\frac{2k+b_{j}}{(k^{2}+b_{j}k+c_{j})^{m}}dk+ {\left(C_{j,m}-\frac{B_{j,m}}{2}\right)\int_{k_{min}}^{k_{max}}
{\frac{dk}{(k^{2}+b_{j}k+c_{j})^{m}}dk}
}}}}}
\label{secondterm}
\end{equation}
}
In turn, the integrals in the first sum of eq. \ref{secondterm} can be easily solved:
\begin{equation}
\begin{array}{ll}
\sum_{j=1}^{j^{*}}{\sum_{m=1}^{m_{j}}{\frac{B_{j,m}}{2}\int_{k_{min}}^{k_{max}}
{
\frac{2k+b_{j}}{(k^{2}+b_{j}k+c_{j})^{m}}dk}}}=
\sum_{j=1}^{j^{*}}{\frac{B_{j,m}}{2}
\ln{\frac{(k_{max}^{2}+b_{j}k_{max}+c_{j})}{(k_{min}^{2}+b_{j}k_{min}+c_{j})}}
}+\\
+
\sum_{j=1|m_{j}>1}^{j^{*}}{\sum_{m=1}^{m_{j}}{\frac{B_{j,m}}{2}
\left(\frac{1}{k_{min}^{2}+b_{j}k_{min}+c_{j}}-\frac{1}{k_{max}^{2}+b_{j}k_{max}+c_{j}}\right)
}}\\
\end{array}
\label{eq1}
\end{equation}
and, again, when taking the limit $k_{max}\rightarrow\infty$, only the logarithmic terms diverge: 
\begin{equation}
\sum_{j=1}^{j^{*}}{\sum_{m=1}^{m_{j}}{\frac{B_{j,m}}{2}\int_{k_{min}}^{k_{max}}
{
\frac{2k+b_{j}}{(k^{2}+b_{j}k+c_{j})^{m}}dk}}}=
\sum_{j=1}^{j^{*}}{\frac{B_{j,m}}{2}
\ln{{(k_{max}^{2})}}
}+ \vartheta=\sum_{j=1}^{j^{*}}{B_{j,m}
\ln{(k_{max})}}+\vartheta
\label{log2}
\end{equation}
Finally, we have to analyze the last integrals in eq. \ref{secondterm}. After the following linear transformation:
\begin{equation}
v=\frac{2k+b_{j}}{\sqrt{4c_{j}-b_{j}^{2}}}
\end{equation}
we can rewrite:
\begin{equation}
{
\left(C_{j,m}-\frac{B_{j,m}}{2}\right)
\int_{k_{min}}^{k_{max}}
{\frac{dk}{(k^{2}+b_{j}k+c_{j})^{m}}}=
\left(C_{j,m}-\frac{B_{j,m}}{2}\right)
\left(c_{j}-\left(\frac{b_{j}}{2}\right)^{2}\right)^{1/2-m}
\int_{\frac{2k_{min}+b_{j}}{\sqrt{4c_{j}-b_{j}^{2}}}}^{\frac{2k_{max}+b_{j}}{\sqrt{4c_{j}-b_{j}^{2}}}}{\frac{1}{(v^{2}+1)^{m}}}dv
}
\end{equation}
and, though it is not possible to write an explicit solution for the integrals of the form $\int1/(v^{2}+1)^{m}$, integrating by parts we get:
\begin{equation}
{
\int{\frac{1}{(v^{2}+1)^{m}}dv}=\frac{v}{(2m-2)(v^{2}+1)^{m-1}}+\frac{2m-3}{2m-2}\int{\frac{dv}{(v^{2}+1)^{m-1}}}
}
\end{equation}
which allows to solve the integrals, yielding the appearance of rational and arctangent terms none of which diverges in the limit $k_{max}\rightarrow\infty$. Thus, only logarithmic terms from equations \ref{log1} and \ref{log2} contribute to the divergence of the initial limit of Eq. \ref{theorem}, that can be finally rewritten as follows:
\begin{equation}
{\left|\lim_{k_{max}\rightarrow \infty}{\int_{k_{min}}^{k_{max}}{  \frac{P(k)}{Q(k)}dk  }  }\right|= 
\ln{(k_{max})}\left|\sum_{i=1}^{i^{*}}{A_{i,1}}+\sum_{j=1}^{j^{*}}{B_{j,m}
}\right|+\vartheta
}
\label{mmm}
\end{equation}
Thus, recalling that $|\vartheta|<<\ln(k_{max})$ stands for finite terms, the only requisite for the limit to diverge is that:
\begin{equation}
{
\left(\sum_{i=1}^{i^{*}}{A_{i,1}}+\sum_{j=1}^{j^{*}}{B_{j,m}}\right)\neq0}
\label{finalcondition}
\end{equation}
which is precisely the coefficient of the monomial of degree equal to $deg(Q)-1$ in the numerator $P'(k)=P(k)/Q_{o}$, as can be easily shown after grouping the partial fractions in Eq. \ref{Parfra} into a single one. Therefore, we have demonstrated the initial statement: the condition for the integral in Eq.\ (\ref{theorembis}) to diverge is that $deg(P')=deg(P)\geq deg(Q)-1$.

\begin{acknowledgments}
The authors thank Maitane Gutierrez for assistance with design software to produce some of the figures. J. S was supported by MINECO through the FPU program. This work has been partially supported by the National Natural Science Foundation of China (NSFC) through Grant 61374169 (to CYX), 
by MINECO through Grant FIS2011-25167 (to YM), Comunidad de Arag\'on (Spain) through a grant to the group FENOL and by the EC FET-Proactive Projects PLEXMATH (grant 317614, to YM) and MULTIPLEX (grant 317532 to JS, SM and YM). 
\end{acknowledgments}


\begin{thebibliography}{99}
\bibitem{note1} We further assume $P(k,l)$ to be stationary, i.e., the composed distribution does not change.
\end{thebibliography}
\end{document}